\documentclass[journal=jctcce,manuscript=article,layout=twocolumn]{achemso}%
\pdfoutput=1
\usepackage{graphicx}
\usepackage[]{hyperref}
\hypersetup{
  colorlinks   = true, 
  urlcolor     = black, 
  linkcolor    = blue, 
  citecolor   = blue 
}
\usepackage{color}
\usepackage{tabularx}
\usepackage{dcolumn}
\usepackage{longtable}
\usepackage{tensor}
\usepackage{placeins}
\usepackage{bm}
\usepackage{amsmath}
\usepackage{amsfonts}
\usepackage{amssymb}
\usepackage{textcomp, gensymb}
\providecommand{\U}[1]{\protect\rule{.1in}{.1in}}

\def\be{\begin{equation}} %
\def\ee{\end{equation}} %
\newcommand{\bea}{\begin{eqnarray}}
\newcommand{\eea}{\end{eqnarray}}
\title{Measurement optimization techniques for excited electronic states in near-term quantum computing algorithms}
\author{Seonghoon Choi}
\affiliation[UTSC]{Department of Physical and Environmental Sciences, University of Toronto Scarborough, Toronto, Ontario M1C 1A4, Canada}
\alsoaffiliation[UofT]{Chemical Physics Theory Group, Department of Chemistry, University of Toronto, Toronto, Ontario M5S 3H6, Canada}
\author{Artur F. Izmaylov}
\email{artur.izmaylov@utoronto.ca}
\affiliation[UTSC]{Department of Physical and Environmental Sciences, University of Toronto Scarborough, Toronto, Ontario M1C 1A4, Canada}
\alsoaffiliation[UofT]{Chemical Physics Theory Group, Department of Chemistry, University of Toronto, Toronto, Ontario M5S 3H6, Canada}
\date{\today}
\begin{document}

\begin{abstract}
The variational quantum eigensolver (VQE) remains one of the most popular near-term quantum algorithms for solving the electronic structure problem. Yet, for its practicality, the main challenge to overcome is improving the quantum measurement efficiency. Numerous quantum measurement techniques have been developed recently, but it is unclear how these state-of-the-art measurement techniques will perform in extensions of VQE for obtaining excited electronic states. Assessing the measurement techniques' performance in the excited state VQE is crucial because the measurement requirements in these extensions are typically much greater than in the ground state VQE, as one must measure the expectation value of multiple observables in addition to that of the electronic Hamiltonian. Here, we adapt various measurement techniques to two widely used excited state VQE algorithms: multi-state contraction and quantum subspace expansion. Then, the measurement requirements of each measurement technique are numerically compared. We find that the best methods for multi-state contraction are ones utilizing Hamiltonian data and wavefunction information to minimize the number of measurements. In contrast, randomized measurement techniques are more appropriate for quantum subspace expansion, with many more observables of vastly different energy scales to measure. Nevertheless, when the best possible measurement technique for each excited state VQE algorithm is considered, significantly fewer measurements are required in multi-state contraction than in quantum subspace expansion.
\end{abstract}
\graphicspath{{./Figures/}}

\section{Introduction}
\label{sec:intro}
The variational quantum eigensolver (VQE)\cite{Peruzzo_OBrien:2014,McClean_Aspuru-Guzik:2016,Rybinkin_Izmaylov:2020,Cerezo_Coles:2021,Anand_Aspuru-Guzik:2022} is the best-known noisy intermediate-scale quantum (NISQ) algorithm\cite{Preskill:2018} for solving the electronic structure problem, i.e., finding the ground state energy of the molecular electronic Hamiltonian,
\begin{equation}
\hat{H}_{e} = \sum_{p,q=1}^{N} h_{pq} \hat{E}_{q}^{p} + \sum_{p,q,r,s=1}^{N} g_{pqrs} \hat{E}_{q}^{p}\hat{E}_{s}^{r}, \label{eq:ferm_ham}
\end{equation}
expressed here in terms of  one-electron excitation operators, $\hat{E}_{q}^{p} = \hat{a}^{\dagger}_{p} \hat{a}_{q}$, and $N$ denotes the number of spin-orbitals in the system. As a hybrid quantum-classical algorithm, VQE finds the ground state energy iteratively: On a quantum computer, the parameterized trial wavefunction $|\psi(\theta)\rangle$ 
is prepared and its Hamiltonian expectation value $E(\theta) = \langle \psi(\theta) | \hat{H}_{e} | \psi(\theta) \rangle$ is measured. Then, using a classical optimizer, the $\theta$ parameters are updated to find the extremum of $E(\theta)$. Thus obtained extremum of $E(\theta)$ corresponds to VQE's solution to the electronic structure problem. Accordingly, measuring the Hamiltonian expectation value is one of the main components of the VQE algorithm. Moreover, this quantum measurement of the Hamiltonian expectation value represents the principal obstacle to the practicality of VQE.\cite{Gonthier_Romero:2022} The challenge in performing an accurate measure of the Hamiltonian expectation value can be illustrated by transforming the electronic Hamiltonian [Eq.~(\ref{eq:ferm_ham})] into its qubit representation,
\begin{equation}
\hat{H}_{q} = \sum_{j=1} c_{j} \hat{P}_{j}, \label{eq:qub_ham}
\end{equation}
where each $N$-qubit Pauli product, $\hat{P}_{j} = \bigotimes_{n=1}^{N} \hat{\sigma}_{n}$, is a tensor product of single-qubit operators: $\hat{\sigma}_{n} \in \{\hat{1}_{n}, \hat{x}_{n}, \hat{y}_{n}, \hat{z}_{n} \}$. Many $\hat{P}_{j}$ terms in $\hat{H}_{q}$ are not all-$\hat{z}$ Pauli products, while digital quantum computers can only measure polynomial functions of Pauli-$\hat{z}$ operators. 

Yet, the expectation value of an arbitrary non-$\hat{z}$ operator, $\hat{A}$, can still be determined via a two-step procedure: First, one finds a unitary operator, $\hat{U}_{A}$, that transforms $\hat{A}$ into its measurable Ising form ($\hat{A}_{I} = \hat{U}_{A} \hat{A} \hat{U}_{A}^{\dagger} = \sum_{i} a_{i} \hat{z}_{i} + \sum_{ij} a_{ij} \hat{z}_{i} \hat{z}_{j} + \dots$). Then, one measures the expectation value of $\hat{A}$ using $\langle \hat{A} \rangle_{\psi} = \langle \hat{A}_{I} \rangle_{\hat{U}_{A} \psi}$, where $\langle \hat{A} \rangle_{\psi} \equiv \langle \psi | \hat{A} | \psi \rangle$. Here and in what follows, we omit $\theta$ for brevity. While there certainly exists a unitary operator $\hat{U}_{H}$ transforming $\hat{H}_{q}$ into its Ising form, finding $\hat{U}_{H}$ would be equivalent to diagonalizing $\hat{H}_{q}$, which is the original hard problem we intend to solve with VQE.

Instead, the majority of commonly employed measurement techniques obtain the Hamiltonian expectation value, $\langle \hat{H} \rangle_{\psi} = \langle \hat{H}_{e} \rangle_{\psi} = \langle \hat{H}_{q} \rangle_{\psi}$, by measuring the expectation value of multiple fragments of the Hamiltonian ($\hat{H}_{\alpha}$). Each $\hat{H}_{\alpha}$ is chosen such that it can be diagonalized by a unitary operator that is easy both to obtain on a classical computer and to implement on a quantum computer ($\hat{U}_{\alpha}$). The total Hamiltonian expectation value can then be obtained as a sum of fragment expectation values: $\langle \hat{H} \rangle_{\psi} = \sum_{\alpha} \langle \hat{H}_{\alpha} \rangle_{\psi}$. Numerous measurement techniques exist under this framework, and the figure of merit often used to quantify their success is the number of measurements required to obtain $\langle \hat{H} \rangle_{\psi}$ with a desired accuracy. As each fragment is measured independently, the total number of measurements required to obtain $\langle \hat{H} \rangle_{\psi}$ with an error below $\epsilon$ is\cite{Crawford_Brierley:2021, Yen_Izmaylov:2023}
\begin{equation}
M(\epsilon) = \frac{1}{\epsilon^{2}} \sum_{\alpha} \frac{\mathrm{Var}_{\psi}(\hat{H}_{\alpha})}{m_{\alpha}}, \label{eq:metric}
\end{equation}
where $\mathrm{Var}_{\psi}(\hat{H}_{\alpha}) = \langle \hat{H}_{\alpha}^{2} \rangle_{\psi} - \langle \hat{H}_{\alpha} \rangle_{\psi}^{2}$ is the variance of $\hat{H}_{\alpha}$, and $m_{\alpha}$ is the fraction of the total number of measurements allocated for measuring $\hat{H}_{\alpha}$.

The $M(\epsilon)$ metric varies significantly depending on the quantum measurement scheme employed to obtain $\hat{H}_{\alpha}$ and $\hat{U}_{\alpha}$. Many methods obtain $\hat{H}_{\alpha}$ and the corresponding $\hat{U}_{\alpha}$ according to an optimized deterministic scheme.\cite{Huggins_Babbush:2021, Yen_Izmaylov:2021, Cohn_Parrish:2021, Choi_Izmaylov:2023, Yen_Izmaylov:2020, Crawford_Brierley:2021, Yen_Izmaylov:2021, Verteletskyi_Izmaylov:2020, Yen_Izmaylov:2023, Choi_Izmaylov:2022, Bonet-Monroig_OBrien:2020, Gresch_Kliesch:2023} Alternatively, one can first sample $\hat{U}_{\alpha}$ probabilistically,\cite{Hadfield_Mezzacapo:2022, Huang_Preskill:2020, Huang_Preskill:2021, Hilmich_Wille:2021, Wu_Yuan:2021, Hadfield:2021, Zhao_Miyake:2021} then compose $\hat{H}_{\alpha}$ by taking a linear combination of every operator that is rotated into the measurable Ising form by $\hat{U}_{\alpha}$. The latter probabilistic approaches are commonly known as classical shadow-based techniques. Typically, if one is interested in measuring the expectation value of a single operator (e.g., $\hat{H}_{e}$), the deterministic methods have lower $M(\epsilon)$'s.\cite{Yen_Izmaylov:2023, Choi_Izmaylov:2022, Choi_Izmaylov:2023} On the other hand, if one wishes to measure the expectation values of many observables simultaneously [e.g., all $k$-body reduced density matrices ($k$-RDM)], the classical shadow-based schemes are superior.\cite{Huang_Preskill:2020, Zhao_Miyake:2021}

Several studies exist that compare the performance of measurement techniques in ground state VQE (e.g., see Refs.~\citenum{Yen_Izmaylov:2021, Yen_Izmaylov:2023, Choi_Izmaylov:2022, Choi_Izmaylov:2023}). Following our expectations, these studies suggest that deterministic measurement schemes that utilize the Hamiltonian ($\hat{H}_{e}$) and quantum state ($|\psi\rangle$) information to minimize $M(\epsilon)$ outperform classical shadow techniques. 

However, it is unclear how effective these optimized measurement techniques would perform in extensions of VQE developed for excited electronic states.\cite{McClean_deJong:2017, Colless_Siddiqi:2018, Takeshita_McClean:2020, McClean_Neven:2020, Yoshioka_Endo:2022, Parrish_Martinez:2019, Parrish_Martinez:2019a, Urbaneck_deJong:2020, Huang_Galli:2022, Tammaro_Motta:2022, Barison_Motta:2022, Lotstedt_Tachikawa:2021, Lotstedt_Tachikawa:2022} In popular excited state VQE algorithms, including quantum subspace expansion (QSE)\cite{McClean_deJong:2017, Colless_Siddiqi:2018, Takeshita_McClean:2020, McClean_Neven:2020, Yoshioka_Endo:2022} and multi-state contraction (MC),\cite{Parrish_Martinez:2019, Parrish_Martinez:2019a} one has to measure many expectation values in addition to $\langle \hat{H}_{e} \rangle_{\psi}$. In QSE, one must additionally measure the expectation values of many dressed Hamiltonians, such as $(\hat{E}^{p}_{q})^{\dagger} \hat{H}_{e} \hat{E}^{r}_{s}$, and in MC-VQE, one needs to measure the expectation value of $\hat{H}_{e}$ in multiple states, $|\psi_{n}(\theta)\rangle$, with each state serving as the parameterized ansatz for the $n$th lowest energy eigenstate. With multiple observables to measure, it is thus no longer clear whether optimized deterministic or classical shadow technique would perform better in these two excited state VQE algorithms. 

In this work, we establish the metrics to quantify the quantum measurement costs of QSE and MC-VQE. We then evaluate them for various quantum measurement techniques to investigate the following questions: (1) Which measurement technique has the lowest required number of measurements for a given excited state VQE algorithm? and (2) Which excited state VQE algorithm is more efficient in terms of quantum measurement requirements? 

Note that while the quantum measurement problem also exists in quantum Krylov methods\cite{Parrish_McMahon:2019, Huggins_Whaley:2020, Motta_Chan:2020, Stair_Evangelista:2020, Cortes_Gray:2022} and methods targeting highly excited states,\cite{Chiew_Kwek:2023} which are promising alternatives to QSE and MC-VQE, consideration of such methods is outside the scope of the current work. Furthermore, the performance analysis exploring the impact of the non-unit quantum gate fidelities and error mitigation techniques will be the topic of our future work. 

\section{Theory}
\label{sec:theory}
\subsection{Estimation of the required number of measurements in QSE and MC-VQE}
\label{subsec:M_req}
To assess the performance of measurement techniques, we must first find analytic expressions for our figure of merit: the number of required measurements in QSE and MC. In QSE, after the normal termination of ground state VQE, one performs the configuration interaction singles (CIS) expansion of the final VQE state, $|\psi\rangle$, to form a linear subspace spanned by $|\psi\rangle$ and $\hat{E}_{q}^{p} |\psi \rangle$. Then, the optimal approximation to the eigenvalues of $\hat{H}_{e}$ within this subspace is found by solving the generalized eigenvalue problem:
\begin{equation}
\mathbf{H} \mathbf{C} = \mathbf{S} \mathbf{C} \mathbf{E}, \label{eq:gen_eigen_prob}
\end{equation}
where 
\begin{equation}
H_{IJ} = \langle \hat{O}_{I}^{\dagger} \hat{H}_{e} \hat{O}_{J} \rangle_{\psi} \label{eq:H_QSE}
\end{equation}
is the subspace Hamiltonian, and
\begin{equation}
S_{IJ} = \langle \hat{O}_{I}^{\dagger} \hat{O}_{J} \rangle_{\psi} \label{eq:S_QSE}
\end{equation}
 is the overlap matrix. For notational simplicity, we have introduced the expansion operators $\hat{O}_{I} \in \{\hat{1}, \hat{E}_{q}^{p} \}$ in Eqs.~(\ref{eq:H_QSE}) and (\ref{eq:S_QSE}). The solution to Eq.~(\ref{eq:gen_eigen_prob}) is obtained as the matrix of eigenvectors, $\mathbf{C}$, and the diagonal matrix of eigenvalues, $\mathbf{E}$. While the generalized eigenvalue problem can be solved efficiently on a classical computer, the $H_{IJ}$ and $S_{IJ}$ matrix elements are measured on a quantum computer.

To measure $H_{IJ}$ and $S_{IJ}$, one can apply any qubit or Majorana algebra-based measurement scheme designed for ground state VQE (see \hyperref[appendixa]{Appendix~A} for details on different measurement schemes) with only a minor modification. For concreteness, we will demonstrate this modification using Pauli products. Nonetheless, what follows is easily transferable to Majorana-based techniques because the mapping between Pauli products and Majorana operators is bijective up to a phase (under JW or BK transformation). 

To start, we represent every operator that must be measured ($\hat{A}_{n} \in \{ \hat{O}_{I}^{\dagger} \hat{H}_{e} \hat{O}_{J}, \hat{O}_{I}^{\dagger} \hat{O}_{J} \}$) in their qubit operator form:
\begin{equation}
\hat{A}_{n} = \sum_{k \in \mathcal{A}_{n}} c_{n,k} \hat{P}_{k}, \quad \mathrm{for}\ n=1, \dots, N_{\mathrm{op}}, \label{eq:A_n}
\end{equation}
where $\mathcal{A}_{n}$ is a set of indices labeling the Pauli products that appear in $\hat{A}_{n}$. Equation~(\ref{eq:A_n}) shows that in QSE, one must measure $N_{\mathrm{op}}$ different linear combinations of $N_{P}$ Pauli products: $\hat{P}_{k}$ for $k \in \bigcup_{n=1}^{N_{\mathrm{op}}} \mathcal{A}_{n} \equiv \{ 1, \dots, N_{P} \}$. Employing one of the measurement techniques designed for ground state VQE, we can (either deterministically or probabilistically) form sets of simultaneously measurable $\hat{P}_{k}$'s. For further details on how we adapt the measurement techniques to find these measurable sets in QSE, see Sec.~\ref{subsec:modification}. Each thus-found set has an associated unitary operator, $\hat{U}_{\alpha}$, that transforms every Pauli product in the set into measurable all-$\hat{z}$ Pauli products. We will use $\mathcal{S}_{\alpha}$ to denote a set of indices labeling the Pauli products in the $\alpha$th measurable set. Note that $\mathcal{S}_{\alpha}$'s may have non-zero intersection. In \hyperref[appendixb]{Appendix~B}, we provide a small example, which serves to clarify our notation: $\mathcal{A}_{n}$ and $\mathcal{S}_{\alpha}$ in particular.

The fragments corresponding to each $\hat{A}_{n}$ operator are determined by $\mathcal{S}_{\alpha}$:
\begin{equation}
\hat{A}_{n} = \sum_{\alpha} \hat{A}_{n}^{(\alpha)} = \sum_{\alpha} \sum_{k \in \mathcal{A}_{n} \cap \mathcal{S}_{\alpha}} c_{n,k}^{(\alpha)} \hat{P}_{k}; \label{eq:QSE_frag}
\end{equation}
the $c_{n,k}^{(\alpha)}$ coefficients are prescribed by the chosen measurement scheme and satisfy $\sum_{\alpha} c_{n,k}^{(\alpha)} = c_{n,k}$. Using Eq.~(\ref{eq:QSE_frag}), one can obtain an unbiased estimate for every $\langle \hat{A}_{n} \rangle_{\psi}$. These $\langle \hat{A}_{n} \rangle_{\psi}$ values are estimated by the classical post-processing of the outcomes obtained from measuring $M \cdot m_{\alpha}$ times each $\alpha$th set of $\hat{P}_{k}$'s (i.e., ones with $k \in \mathcal{S}_{\alpha}$). The estimator for $\langle \hat{A}_{n} \rangle_{\psi}$ is the sum of the estimators for each $\langle \hat{A}_{n}^{(\alpha)} \rangle_{\psi}$, i.e., $\bar{A}_{n} = \sum_{\alpha} \bar{A}_{n}^{(\alpha)}$, and the $\hat{A}_{n}^{(\alpha)}$ fragment estimator ($\bar{A}_{n}^{(\alpha)}$) is obtained by averaging $M \cdot m_{\alpha}$ measurement outcomes: 
\begin{equation}
\bar{A}_{n}^{(\alpha)} =\sum_{k \in \mathcal{A}_{n} \cap \mathcal{S}_{\alpha}} c_{n,k}^{(\alpha)} \left[ \frac{1}{M \cdot m_{\alpha}}  \sum_{i=1}^{M \cdot m_{\alpha}} P_{k}^{(i, \alpha)} \right], \label{eq:sample_average}
\end{equation}
where $P_{k}^{(i, \alpha)}$ is the measurement outcome of $\hat{U}_{\alpha} \hat{P}_{k} \hat{U}_{\alpha}^{\dagger}$ from the $i$th collapse of $\hat{U}_{\alpha}|\psi\rangle$.

To quantify the performance of a measurement scheme for QSE, we compute the number of measurements required to provide an upper bound for the error of every $\bar{A}_{n}$ estimator by $\epsilon$, i.e., to satisfy $[\mathrm{Var}(\bar{A}_{n})]^{1/2} \leq \epsilon$ for all $n = 1, \dots, N_{\mathrm{op}}$. While this does not guarantee that the errors in the final eigenvalues obtained by solving the generalized eigenvalue equation [Eq.~(\ref{eq:gen_eigen_prob})] are also going to be below $\epsilon$, the analysis based on first-order perturbation theory\cite{Yoshioka_Endo:2022} and a more rigorous analysis involving the thresholding technique\cite{Epperly_Nakatsukasa:2022} show that the eigenvalue errors are upper bounded approximately by $\mathcal{O}(c \sqrt{D} \cdot \epsilon)$, where $c$ the condition number associated with the eigenvalue problem, and $D$ is the dimension of the $\mathbf{H}$ matrix.\cite{Mathias_Li:2004, Vershynin:2018, Epperly_Nakatsukasa:2022}

Because the Pauli products in $\hat{A}_{n}^{(\alpha)}$ fragments having different $\alpha$'s are measured independently, the covariance between the different fragment estimators is zero: $\mathrm{Cov}(\bar{A}_{n}^{(\alpha)}, \bar{A}_{n}^{(\beta)})$, $\alpha \neq \beta$. Therefore, the total estimator ($\bar{A}_{n}$) variance is simply the sum of fragment estimator variances: $\mathrm{Var}(\bar{A}_{n}) = \sum_{\alpha} \mathrm{Var}(\bar{A}_{n}^{(\alpha)})$. Central limit theorem states that $\bar{A}_{n}^{(\alpha)}$, the sample average of $M \cdot m_{\alpha}$ measurement outcomes [see Eq.~(\ref{eq:sample_average})], follows a normal distribution with mean $\mu_{\bar{A}_{n}^{(\alpha)}} = \langle \hat{A}_{n}^{(\alpha)} \rangle_{\psi}$ and variance $\mathrm{Var}(\bar{A}_{n}^{(\alpha)}) = \mathrm{Var}_{\psi}(\hat{A}_{n}^{(\alpha)})/(M \cdot m_{\alpha})$ for large $M \cdot m_{\alpha}$. Therefore,
\begin{equation}
| \langle \hat{A}_{n} \rangle_{\psi} - \bar{A}_{n} | \leq \sqrt{\sum_{\alpha} \frac{\mathrm{Var}_{\psi}(\hat{A}_{n}^{(\alpha)})}{M \cdot m_{\alpha}}}  \label{eq:var_bar_A_n}
\end{equation}
holds with a 68.27\% chance (for a higher confidence, one can multiply the estimate by $s > 1$, e.g., one can be 99.73\% sure if $s=3$).
Although the central limit theorem is formally an approximation, in our examples, it is very well justified because $M$ is usually very large ($> 10^{6}$), and therefore $M \cdot m_{\alpha}$ is also very large.
By re-arranging Eq.~(\ref{eq:var_bar_A_n}), we obtain the number of measurements required to estimate all $\langle \hat{A}_{n} \rangle_{\psi}$ expectation values with an error $\leq \epsilon$:
\begin{equation}
M(\epsilon) = \frac{1}{\epsilon^{2}} \max_{n} \left[\sum_{\alpha} \frac{1}{m_{\alpha}} \mathrm{Var}_{\psi}(\hat{A}_{n}^{(\alpha)}) \right], \label{eq:QSE_metric}
\end{equation}
which is our figure of merit for QSE. To avoid limiting our assessment to a specific type of VQE ansatz, we assume $| \psi\rangle = | \mathrm{FCI} \rangle$. We acknowledge that while the aim of VQE is to converge to a state close to $| \mathrm{FCI} \rangle$, some VQE ansatzes may not successfully converge to $| \psi\rangle \approx | \mathrm{FCI} \rangle$.

In MC-VQE, the VQE iterations themselves are modified to account for the excited states. Unlike in the ground state VQE, where $\theta$ values parameterize a single state [i.e., $|\psi(\theta)\rangle \equiv \hat{U}(\theta) | \varphi_{0} \rangle$], $\theta$'s parameterize multiple states in MC-VQE. Therefore, the ansatz can be expressed as $|\psi_{n} (\theta)\rangle = \hat{U}(\theta) | \varphi_{n} \rangle$, where $|\varphi_{n}\rangle$'s are classically efficient orthonormal reference states (e.g., CIS states), and $\hat{U}(\theta)$ are parameterized entanglers implemented on a quantum circuit. The hybrid quantum-classical iterative procedure (like the one in the ground state VQE) is then used to optimize the state-averaged energy, $E_{\mathrm{avg}}(\theta) \propto \sum_{n} \langle \hat{H}_{e} \rangle_{\psi_{n}(\theta)}$. 

In contrast to QSE, where the one has to measure the expectation values of many $\hat{A}_{n}$ observables in a single state, in MC-VQE, one must measure the expectation value of a single $\hat{H}_{e}$ operator in multiple states [$|\psi_{n} (\theta) \rangle$]. The metric for the quantum measurement cost in MC-VQE is thus different from Eq.~(\ref{eq:QSE_metric}). The number of measurements required in MC-VQE can be estimated by summing the number of measurements required to obtain each $\langle \hat{H}_{e} \rangle_{\psi_{n}(\theta)}$ expectation value with $\epsilon$ accuracy. As in the metric used for QSE, to prevent restricting our study to a specific VQE ansatz type, we use the exact ground and excited states, obtained by diagonalizing $\hat{H}_{e}$, to estimate the quantum measurement cost, and hence our figure of merit becomes
\begin{equation}
M(\epsilon) = \sum_{n}\left[ \frac{1}{\epsilon^{2}} \sum_{\alpha} \frac{\mathrm{Var}_{\psi_{n}} (\hat{H}_{\alpha})}{m_{\alpha}} \right], \label{eq:MC_metric}
\end{equation}
where $\psi_{n}$ are the exact ground and excited states. Note that the required number of measurements for a single state is given by Eq.~(\ref{eq:metric}).\cite{Crawford_Brierley:2021, Yen_Izmaylov:2023} 

In Eq.~(\ref{eq:MC_metric}), we chose to include only a small number of lowest energy states for two reasons: (1) many chemically motivated applications only require the lowest few energy levels, and (2) the QSE algorithm limited to CIS expansions can only accurately obtain the lowest few eigenvalues of the molecular electronic Hamiltonian.\cite{McClean_deJong:2017} In our assessment, we chose to include the 10 lowest lying states for computing the MC-VQE metric. Although our choice is somewhat arbitrary, it can be warranted considering the main focus of our study. Our aim is to figure out whether the optimization in the deterministic measurement schemes would still be sufficiently efficient to outperform shadow-based techniques when one has to measure the expectation value in multiple states that are close in energy.

\subsection{Adapting the measurement schemes for the excited state VQE algorithms}
\label{subsec:modification}
Because most measurement schemes have been developed specifically for ground state VQE, small modifications are necessary to make them applicable to the excited state VQE algorithms. For a summary of common quantum measurement techniques, with a focus on those employed in this work, we refer the reader to \hyperref[appendixa]{Appendix~A}. In this study, we consider the following popular measurement schemes: fully commuting (FC), qubit-wise commuting (QWC), and Majorana classical shadow (CS),\cite{Huang_Preskill:2020, Zhao_Miyake:2021} the derandomized extension of QWC classical shadow (Derand),\cite{Huang_Preskill:2021} FC and QWC sorted insertion (SI) algorithm,\cite{Crawford_Brierley:2021} FC and QWC iterative measurement allocation (IMA),\cite{Yen_Izmaylov:2023} FC iterative coefficient splitting (ICS),\cite{Yen_Izmaylov:2023} and fluid fermionic fragments (F$^{3}$).\cite{Choi_Izmaylov:2023} Specifically, the ``Full'' version\cite{Choi_Izmaylov:2023} of F$^{3}$ based on the low-rank (LR)\cite{Berry_Babbush:2019, Motta_Chan:2021, Huggins_Babbush:2021} decomposition was chosen due to its efficient combination of low classical computational cost with reduced quantum measurements.  Note that our list of measurement techniques is not exhaustive but still includes the best-performing representatives within each class of measurement techniques.\cite{Crawford_Brierley:2021, Yen_Izmaylov:2023, Huang_Preskill:2021, Zhao_Miyake:2021} 

Among the considered methods, FC-ICS and F$^{3}$ will only be employed for MC-VQE because they are not straightforwardly adaptable to QSE. The classical cost of FC-ICS and F$^3$ would be prohibitive due to the $\mathrm{max}_{n}$ function in the QSE metric [Eq.~(\ref{eq:QSE_metric})] and the large number of $\hat{A}_{n}$ operators to measure. Moreover, LR decomposition,\cite{Berry_Babbush:2019, Motta_Chan:2021, Huggins_Babbush:2021} on which F$^3$ is based, has not been developed for dressed Hamiltonians ($\hat{O}_{I}^{\dagger} \hat{H}_{e} \hat{O}_{J}$) appearing in QSE. We also do not include POVM-based measurement optimization\cite{Garcia-Perez_Maniscalco:2021, Glos_Garcia-Perez:2022} because it is unclear how such techniques can be applied to either of the excited state VQE algorithms we consider.

The qubit-based (FC and QWC) CS and Majorana-CS methods, which exploit neither the operator nor the state information, are directly applicable to both QSE and MC-VQE without any modification: In these CS methods, the $\hat{U}_{\alpha}$ unitaries fixing the measurement frames are drawn from a probability distribution independent of the state and the operators to be measured. 

In contrast, the Derand extension determines $\hat{U}_{\alpha}$'s based on the importance of the Pauli product terms to measure. In MC-VQE, we have a single observable ($\hat{H}_{q} = \sum_{j} c_{j} \hat{P}_{j}$) to measure, and the importance of $\hat{P}_{j}$ is ranked by the $| c_{j} |$ magnitudes, as in the Derand for ground state VQE.\cite{Huang_Preskill:2021} On the other hand, in QSE, one must account for all $\hat{A}_{n} = \sum_{k \in \mathcal{A}_{n}} c_{n,k} \hat{P}_{k}$ operators that must be measured. We chose to use the sum of magnitudes, $\sum_{n:k \in \mathcal{A}_{n}} |c_{n,k}|$, to quantify the importance of $\hat{P}_{k}$ when choosing $\hat{U}_{\alpha}$ in Derand. Likewise, in SI, IMA, and ICS, the Pauli products contained in each simultaneously measured set are determined based on the magnitude of their coefficients through a greedy algorithm.\cite{Crawford_Brierley:2021, Yen_Izmaylov:2023} The same quantities used in Derand ($|c_{j}|$ for MC-VQE and $\sum_{n:k \in \mathcal{A}_{n}} |c_{n,k}|$ for QSE) was also used for this greedy algorithm.

On top of the operator information, the optimized deterministic measurement schemes (SI, IMA, ICS, and F$^{3}$) also use the information about the quantum state to lower $M(\epsilon)$. The metrics used for the excited state VQEs [Eq.~(\ref{eq:QSE_metric}) for QSE and Eq.~(\ref{eq:MC_metric}) for MC-VQE] are different from that for the ground state VQE [Eq.~(\ref{eq:metric})], and therefore the cost function in the optimization has to be modified accordingly.

In QSE, the aim is to minimize Eq.~(\ref{eq:QSE_metric}), but because the information about the quantum state ($|\psi \rangle$) is often not classically available, the CISD states are used instead as the classically efficient proxy. Moreover, since in QSE, the minimization of $M(\epsilon)$ with respect to either $m_{\alpha}$ or $\hat{A}_{n}^{(\alpha)}$ is non-linear due to the $\max_{n}$ function, we make a further approximation and minimize instead
\begin{equation}
M_{\phi}(\epsilon) = \frac{1}{\epsilon^{2}} \sum_{\alpha} \frac{1}{m_{\alpha}} \max_{n}[\mathrm{Var}_{\phi}(\hat{A}_{n}^{(\alpha)})], \label{eq:M_phi_QSE}
\end{equation}
where $|\phi\rangle$ is the CISD ground state. 

In MC-VQE, to uniformly account for the variances of every state of interest ($|\psi_{n}\rangle$ for $n=1,\dots,N_{s}$), we use the variances of the mixed CISD state, $\rho = (\sum_{n=1}^{N_{s}} |\phi_{n}\rangle\langle\phi_{n}|) / \sqrt{N_{s}}$, as a proxy for the average variances, [$\sum_{n=1}^{N_{s}} \mathrm{Var}_{\psi_{n}}(\hat{H}_{\alpha})]/N_{s}$. In this study, $N_{s}=10$. The cost function in the measurement optimization for MC-VQE thus becomes
\begin{equation}
M_{\rho}(\epsilon) = \frac{1}{\epsilon^{2}} \sum_{\alpha}  \frac{\mathrm{Var}_{\rho} (\hat{H}_{\alpha})}{m_{\alpha}}, \label{eq:M_rho}
\end{equation}
where $\mathrm{Var}_{\rho}(\hat{H}_{\alpha}) = \mathrm{Tr}(\hat{H}_{\alpha}^{2} \rho) - \mathrm{Tr}(\hat{H}_{\alpha} \rho)^{2}$.

\section{Results and discussion}
\label{sec:results}
We numerically compare the figure of merit, $\epsilon^{2} M(\epsilon)$, in various measurement schemes applied to QSE and MC-VQE. In QSE, this figure of merit corresponds to the number of measurements in millions required to obtain all QSE matrix elements [Eqs.~(\ref{eq:H_QSE}) and (\ref{eq:S_QSE})] with better than or equal to $10^{-3}$ a.u. accuracy. In MC-VQE, the $\epsilon^{2} M(\epsilon)$ value indicates the number of measurements in millions to measure all $\langle \hat{H}_{e} \rangle_{\psi_{n}}$ with $n \le N_{s}$ with $10^{-3}$ a.u. accuracy ($N_{s}=10$ with an exception of H$_2$ in STO-3G basis where only 6 states exist). 

We apply the different measurement schemes to compute $\epsilon^{2} M(\epsilon)$ for electronic Hamiltonians of several molecules in the STO-3G basis and the following nuclear geometries: $R(\mathrm{H} - \mathrm{H}) = 1$\AA\ with $\angle\mathrm{H}\mathrm{H}\mathrm{H}=180\degree$ (for H$_{3}^{+}$, H$_{4}$, and H$_{6}$), $R(\mathrm{H} - \mathrm{X}) = 1$\AA\ (for X$=$H, F, Li), $R(\mathrm{Be} - \mathrm{H}) = 1$\AA\ with $\angle\mathrm{H}\mathrm{Be}\mathrm{H}=180\degree$ (for BeH$_{2}$), $R(\mathrm{O} - \mathrm{H}) = 1$\AA\ with $\angle\mathrm{H}\mathrm{O}\mathrm{H}=107.6\degree$ (for H$_{2}$O). The data for these molecular electronic Hamiltonians [i.e., $h_{pq}$ and $g_{pqrs}$ in Eq.~(\ref{eq:ferm_ham})] are available in Ref.~\citenum{Choi_Izmaylov_data:2022}. For the deterministic techniques based on qubit algebra (SI,\cite{Crawford_Brierley:2021} IMA,\cite{Yen_Izmaylov:2023} ICS\cite{Yen_Izmaylov:2023}), we only consider the JW transformation because it was shown in Ref.~\citenum{Choi_Izmaylov:2022} that no appreciable differences exist between the $\epsilon^{2} M(\epsilon)$ values for $\hat{H}_{q}$ obtained with JW and BK. As for the shadow-based techniques using qubit operator algebra,\cite{Huang_Preskill:2020, Huang_Preskill:2021} we consider both JW and BK transformations but only report the results with a lower $\epsilon^{2} M(\epsilon)$ value for each system for brevity.

\begin{table*}[h!]
\caption{\label{tab:QSE_extended} Number of measurements in millions required to obtain every QSE matrix element with $10^{-3}$ a.u. accuracy [$\epsilon^{2} M(\epsilon)$] for systems with $N \leq 8$. Only the fermion-qubit mapping with a lower $\epsilon^{2} M(\epsilon)$ is shown for qubit shadow-based techniques (indicated in parentheses).}
\resizebox{\textwidth}{!}{
\begin{tabular}{*9{c}}
\hline
\hline
Sys         & QWC-SI & QWC-IMA& FC-SI & FC-IMA& QWC-CS    & FC-CS      & Derand      & Majorana-CS \\
\hline                                                                              
H$_{2}$     & 6.25   & 5.68   & 2.82  & 2.82  & 7.06 (BK) & 7.59 (BK)  & 5.36 (BK)   & 2.11        \\
H$_{3}^{+}$ & 27.0   & 16.0   & 6.42  & 6.42  & 102 (BK)  & 29.7 (JW)  & 19.1 (BK)   & 3.09      \\
H$_{4}$     & 177    & 81.6   & 72.9  & 54.2  & 289 (BK)  & 124 (JW)   & 163 (BK)    & 18.8    \\
\hline
\hline
\end{tabular}
}
\end{table*}

We will first compare the different measurement techniques for each excited state VQE. Then, we will determine whether QSE or MC-VQE is more practical in regards to quantum measurement by considering only the best measurement technique for each excited state VQE. For brevity, our discussion will be focused on the best-performing variant out of the following classes of techniques: (1) deterministic qubit algebra-based methods: SI,\cite{Crawford_Brierley:2021} IMA,\cite{Yen_Izmaylov:2023} and ICS;\cite{Yen_Izmaylov:2023} (2) shadow-based techniques: Majorana CS,\cite{Zhao_Miyake:2021} qubit CS,\cite{Huang_Preskill:2020} and Derand;\cite{Huang_Preskill:2021} and (3) F$^{3}$.\cite{Choi_Izmaylov:2023} We remind the reader that ICS and F$^{3}$ are only applicable to MC-VQE. 

Due to the prohibitive classical computational cost of evaluating the $\epsilon^{2} M(\epsilon)$ metric in QSE, we have determined the best method in each class only using results from small systems (with $N \leq 8$). FC-IMA had the lowest $\epsilon^{2} M(\epsilon)$ among the deterministic methods (see Table~\ref{tab:QSE_extended}). However, for larger systems ($N>8$), FC-SI was the best viable deterministic method because the classical computational cost of IMA became prohibitive. The shadow-based technique with the lowest $\epsilon^{2} M(\epsilon)$ was Majorana-CS for QSE. Indeed, Zhao \textit{et al.} have demonstrated that this variant of CS outperforms the qubit counterparts for measuring fermionic $k$-RDMs as $k$ increases.\cite{Zhao_Miyake:2021} Since the $\hat{A}_{n}$ operators (to be measured in QSE) contain up to $4$-RDMs, it can be expected that Majorana-CS outperforms qubit-CS. Moreover, Derand has higher $\epsilon^{2}M(\epsilon)$'s than Majorana-CS because the procedure in Derand optimizing $\hat{U}_{\alpha}$ probability distribution was mainly designed for measuring a single observable (e.g., $\hat{H}_{q}$) accurately. In QSE, one is instead interested in accurately measuring many observables simultaneously.

\begin{table}[h!]
\caption{\label{tab:QSE} Number of measurements in millions required to obtain every QSE matrix element with $10^{-3}$ a.u. accuracy [$\epsilon^{2} M(\epsilon)$]. The $\epsilon^{2} M(\epsilon)$ values in the best methods, FC-SI and Majorana-CS, are compared for Hamiltonians of several molecules ($N$ is the number of spin-orbitals and is equal to the number of qubits, and $N_{\mathrm{op}}$ is the number of $\hat{A}_{n}$ operators to measure in QSE; $N_{P}$ is the total number of Pauli products to measure).}
\resizebox{\columnwidth}{!}{
\begin{tabular}{*6{c}}
\hline
\hline
Sys         & $N$     & $N_{\mathrm{op}}$ & $N_{P}$    & FC-SI      & Majorana-CS    \\ 
\hline                                                                            
H$_{2}$     & 4       & 30                 & 127       & 2.82       & 2.11           \\
H$_{3}^{+}$ & 6       & 90                 & 1879      & 6.42       & 3.09           \\
H$_{4}$     & 8       & 306                & 22060     & 72.9       & 18.8           \\
HF          & 12      & 462                & 269327    & 120000     & 93300          \\
LiH         & 12      & 1122               & 552490    & 2110       & 494            \\
H$_{6}$     & 12      & 1406               & 754771    & 893        & 116            \\
BeH$_{2}$   & 14      & 2450               & 1972969   & -          & $\geq 3000$    \\
H$_{2}$O    & 14      & 1722               & 1810651   & -          & $\geq 77300$   \\
\hline
\hline
\end{tabular}
}
\end{table}

\begin{table*}[h!]
\caption{\label{tab:MC} Number of measurements in millions required to obtain every $\langle \hat{H}_{e} \rangle_{\psi_{n}}$ for $n = 1, \dots\, 10$ with $10^{-3}$ a.u. accuracy [$\epsilon^{2} M(\epsilon)$]. The best method in each class of measurement methods is highlighted in bold. For most systems, the qubit shadow-based techniques performed better under JW transformation. A few exceptions where BK yields a lower $\epsilon^{2} M(\epsilon)$ are marked with an asterisk. For compactness, QWC, FC, and Majorana are shortened to Q, F, and M, respectively.}
\resizebox{\textwidth}{!}{
\begin{tabular}{ccc*9{c}}
\hline
\hline
Sys     & $N_{P}$ & Q-SI & Q-IMA & F-SI & F-IMA & \textbf{F-ICS}  & Q-CS      & F-CS     & \textbf{Derand}     & M-CS & \textbf{F$^{3}$}  \\
\hline                                                                                                     
H$_{2}$    & 14      & 0.502  & 0.502   & 0.340 & 0.340  & \textbf{0.305}   & 3.53   & 2.33 & \textbf{0.494} & 3.22              & \textbf{0.278}      \\
H$_{3}^{+}$& 61      & 6.72   & 6.67    & 2.85  & 2.70   & \textbf{1.90}    & 68.0   & 34.7 & \textbf{9.67$^{\ast}$}  & 46.8     & \textbf{1.04}   \\
H$_{4}$    & 184     & 39.9   & 34.2    & 10.7  & 10.7   & \textbf{9.25}    & 441    & 242$^{\ast}$  & \textbf{41.2}  & 179      & \textbf{3.90}  \\
HF         & 630     & 125    & 69.8    & 55.0  & 55.0   & \textbf{7.20}    & 24400$^{\ast}$  & 10300& \textbf{513}   & 4400     & \textbf{2.35}   \\
LiH        & 630     & 14.6   & 10.3    & 10.5  & 10.5   & \textbf{2.98}    & 782$^{\ast}$    & 583  & \textbf{23.8}  & 394      & \textbf{0.944}   \\
H$_{6}$    & 918     & 287    & 229     & 46.0  & 34.5   & \textbf{21.4}    & 2350   & 3410 & \textbf{269}   & 765               & \textbf{10.7}   \\
BeH$_{2}$  & 665     & 70.2   & 58.6    & 17.6  & 17.6   & \textbf{11.6}    & 2360   & 2150 & \textbf{108}   & 1270              & \textbf{6.80}  \\
H$_{2}$O   & 1085    & 247    & 154     & 77.9  & 77.9   & \textbf{27.3}    & 30300$^{\ast}$  & 24000& \textbf{901}   & 3660     & \textbf{8.28}  \\
\hline
\hline
\end{tabular}
}
\end{table*}
Majorana-CS also has a lower $\epsilon^{2} M(\epsilon)$ than that in the best deterministic measurement technique for QSE: FC-SI. Table~\ref{tab:QSE} shows that the number of required measurements for QSE would be $3.4$ times larger on average if FC-SI is employed instead of Majorana-CS. In particular, the $\epsilon^{2}M(\epsilon)$ value in FC-SI is $7.7$ times higher than that in Majorana-CS for H$_{6}$. The underperformance of FC-SI can be explained by the failure to suitably optimize the measurement allocation [$m_{\alpha}$ in Eq.~(\ref{eq:QSE_metric})]. In FC-SI, $M(\epsilon)$ is lowered by minimizing a classically efficient proxy $M_{\phi}(\epsilon)$ [Eq.~(\ref{eq:M_phi_QSE})]. While it was shown in Refs.~\citenum{Yen_Izmaylov:2023, Choi_Izmaylov:2022, Choi_Izmaylov:2023} that it is a good approximation to use the CISD states to estimate quantum variances, the approximation to move the $\max_{n}$ operation inside the sum over $\alpha$ is a poor one, leading to the failure of FC-SI. This approximation is inaccurate due to the large discrepancy between the values of $\mathrm{Var}_{\psi}(\hat{A}_{n}^{(\alpha)})$ for different $n$. In QSE, one is required to measure $\langle \hat{O}^{\dagger} \hat{H}_{e} \hat{O} \rangle_{\psi}$ for both $\hat{O} = \hat{1}$ and $\hat{O} = \hat{E}_{1}^{N}$, and the variances corresponding to $\hat{H}_{e}$ is expected to be significantly different from those corresponding to $(\hat{E}_{1}^{N})^{\dagger} \hat{H}_{e} \hat{E}_{1}^{N}$ due to the large discrepancy in the energy scales of $\langle \hat{H}_{e} \rangle_{\psi}$ and $\langle (\hat{E}_{1}^{N})^{\dagger} \hat{H}_{e} \hat{E}_{1}^{N} \rangle_{\psi}$. 

When the QSE operators, $\hat{O}_{I}$, are chosen as the CIS expansions, the number of QSE matrix elements ($N_{\mathrm{op}}$) has approximately cubic scaling with system size, $N$;\cite{Sherrill:1996} in Table~\ref{tab:QSE}, $N_{\mathrm{op}}$ has $N^{3.3}$ scaling. Because the number of Pauli products in each QSE matrix element scales as $N^{4}$, the total number of Pauli products to measure in QSE is expected to scale as $\sim N^{7}$. This scaling is supported by our numerical results, which show that $N_{P}$ has $N^{7.7}$ scaling. Moreover, the $N^{7}$ scaling of $N_{P}$ is also consistent with the scaling of $\epsilon^{2} M(\epsilon)$ with $N$. Table~\ref{tab:QSE} shows that $\epsilon^{2} M(\epsilon)$ has $N^{7.7}$ scaling in both Majorana-CS and FC-SI. Note that the classical computational resources required to evaluate $\epsilon^{2} M(\epsilon)$ in large ($N = 14$) molecules were becoming prohibitive for QSE. To be more frugal with the computational resources, we only computed $\epsilon^{2} M(\epsilon)$ for Majorana-CS, the best-performing method for QSE. Moreover, instead of computing $M_{n} = \sum_{\alpha} \mathrm{Var}_{\psi}(\hat{A}_{n}^{(\alpha)})/m_{\alpha}$ for every $n$, we estimate $M(\epsilon) = (\max_{n} M_{n})/\epsilon^{2}$ by evaluating the $M_{n}$ values for only $\sim 75\%$ of $\hat{A}_{n}$ operators. The true $\epsilon^{2} M(\epsilon)$ values for BeH$_{2}$ and H$_{2}$O are lower bounded by the values reported in Table~\ref{tab:QSE}.

In MC-VQE, Derand achieves a marked reduction in $\epsilon^{2} M(\epsilon)$ compared to the standard CS techniques (see Table~\ref{tab:MC}). In contrast to QSE, Derand performs well for MC-VQE because one is interested in measuring a single $\hat{H}_{q}$ observable. Therefore, Derand's optimization of measurement frames, $\hat{U}_{\alpha}$, remains effective. However, even Derand has much higher $\epsilon^{2} M(\epsilon)$ values than either FC-ICS or F$^{3}$. On average, $\epsilon^{2} M(\epsilon)$ in Derand is $18$ times higher than that in FC-ICS and $52$ times higher than that in F$^{3}$. Such comparison is not entirely impartial since $\hat{U}_{\alpha}$'s in FC-ICS and F$^{3}$ require two-qubit gates to implement, whereas $\hat{U}_{\alpha}$'s in Derand can be implemented from single-qubit Clifford gates. Yet, $\epsilon^{2} M(\epsilon)$'s in Derand are still higher than those in QWC-IMA for every molecule except H$_{2}$. Table~\ref{tab:MC} shows on average, $\epsilon^{2} M(\epsilon)$ in Derand was a factor of 2.8 times larger than that in QWC-IMA. In MC-VQE, the deterministic optimization schemes (IMA, ICS, and F$^{3}$) perform well because, unlike in QSE, the energy scales of the measured expectation values ($\langle \hat{H}_{e} \rangle_{\psi_{n}}$ for $n=1, \dots, N_{s}$) are similar. Therefore, optimizing $M_{\rho}(\epsilon)$ [Eq.~(\ref{eq:M_rho})] finds $\hat{H}_{\alpha}$ fragments with a low $\mathrm{Var}_{\psi_{n}}(\hat{H}_{\alpha})$ for all $n \leq N_{s}$.

\begin{table}[h!]
\caption{\label{tab:iter} Number of iterations at which the quantum measurement cost of MC-VQE exceeds that of QSE ($N_{\mathrm{crit}}$).}
\begin{tabular}{*3{c}}
\hline
\hline
Sys      & $N$     & $N_{\mathrm{crit}}$         \\
\hline                    
H$_{2}$     & 4       & 8                 \\
H$_{3}^{+}$ & 6       & 3                  \\
H$_{4}$     & 8       & 5                  \\
HF          & 12      & 39703  \\
LiH         & 12      & 524   \\
H$_{6}$     & 12      & 11                 \\
BeH$_{2}$   & 14      & 442                \\
H$_{2}$O    & 14      & 9340               \\
\hline
\hline
\end{tabular}
\end{table}

While the $\epsilon^{2} M(\epsilon)$ values in Table~\ref{tab:QSE} represent the number of measurements required in QSE in addition to the normal termination of VQE, the values in Table~\ref{tab:MC} show the number of measurements required for every MC-VQE iteration. Therefore, the extra quantum measurement cost incurred by MC-VQE depends on the number of iterations required for convergence. To estimate this extra cost per iteration, we subtract the required number of measurements in a ground state VQE iteration [i.e., $\epsilon^{2} M(\epsilon)$ values of Ref.~\citenum{Choi_Izmaylov:2023}] from the results in Table~\ref{tab:MC}. In Table~\ref{tab:iter}, we present the number of MC-VQE iterations beyond which the measurement cost of MC-VQE becomes higher than that of QSE ($N_{\mathrm{crit}}$). For all small molecules with $N \leq 8$, the $N_{\mathrm{crit}}$ values are smaller than the typically required number of iterations in VQE. However, for all larger systems except H$_6$, $N_{\mathrm{crit}}$ far exceeds the number of VQE iterations typically required for convergence ($\sim 50$). The large $N_{\mathrm{crit}}$ for systems with $N \ge 12$ can be explained by the $\sim N^{3}$ scaling of the number of QSE matrix elements. Furthermore, it can be expected that as $N$ increases, $N_{\mathrm{crit}}$ will also increase due to this $\sim N^{3}$ scaling in QSE; in contrast, the number of observables to measure remains constant in MC-VQE.

\section{Conclusion}
In this work, we have compared several state-of-the-art measurement techniques to assess their ability to lower the number of quantum measurements required in the excited state extensions of VQE. We made the necessary modifications and applied these measurement methods to two different VQE-based algorithms for excited states: quantum subspace expansion (QSE)\cite{McClean_deJong:2017, Colless_Siddiqi:2018, Takeshita_McClean:2020, McClean_Neven:2020, Yoshioka_Endo:2022} and multi-state contraction (MC).\cite{Parrish_Martinez:2019, Parrish_Martinez:2019a} Our numerical analysis showed that in MC-VQE, the fluid fermionic fragment (F$^{3}$) technique, which exploits the classically available information about the electronic Hamiltonian and the wavefunction, yields the lowest number of required measurements. This F$^{3}$ technique was also shown to be most successful in ground state VQE.\cite{Choi_Izmaylov:2023} However, in QSE, the randomized classical shadow technique based on Majorana operator algebra performs best. 

This difference in performance can be explained by the nature of the observables that must be measured in the two different excited state VQE algorithms. In MC-VQE, one measures the expectation value of the Hamiltonian, $\hat{H}_{e}$, in several low-lying states, $|\psi_{n}\rangle$. Because the multiple expectation values that must be measured have similar energies, optimization based on the mixed state, $\rho = (\sum_{n=1}^{N_s} |\psi_{n}\rangle \langle \psi_{n}|) / \sqrt{N_s} $, still yields a low required measurement number for every $\langle \psi_{n} | \hat{H}_{e} | \psi_{n}\rangle$. In QSE, on the other hand, one has to measure a large number of observables with very different energy scales, thereby making optimization challenging. Therefore, the classical shadow technique, agnostic of the Hamiltonian and wavefunction information but equipped with worst-case performance guarantees, performs best in QSE.

The measurement technique with the lowest number of required measurements was different in MC-VQE and QSE. Yet, when the best possible measurement technique was applied for each excited state VQE algorithm, the quantum measurement resource requirement for QSE was much greater than that for MC-VQE. Moreover, while in MC-VQE, the number of measurements required to obtain the excited state energies was larger by approximately a constant factor compared to that required in the ground state VQE, the additional measurement number required for QSE increased rapidly with system size. The more recently developed quantum self-consistent equation-of-motion (q-sc-EOM) method\cite{Asthana_Mayhall:2023} was shown to be more efficient in terms of quantum measurement cost than QSE, but it still suffers from the number of observables to measure scaling as $N^{3}$ with system size. Consequently, we conclude that MC-VQE is the more practical excited state VQE method in terms of measurement requirements.

\section*{Acknowledgments}
S.C. thanks Tzu-Ching Yen for helpful discussions and acknowledges financial support from the Swiss National Science Foundation through the Postdoc Mobility Fellowship (Grant No. P500PN-206649). A.F.I. is grateful for financial support from Zapata Computing Inc. and the Natural Sciences and Engineering Research Council of Canada. This research was enabled in part by support provided by Compute Ontario and Compute Canada.

\section*{Appendix A: Quantum measurement schemes}
\label{appendixa}
The different measurement schemes can largely be classified by the type of algebra they rely upon to find the $\hat{H}_{\alpha}$ fragments. In fermionic operator algebra-based methods,\cite{Berry_Babbush:2019, Motta_Chan:2021, Huggins_Babbush:2021, Yen_Izmaylov:2021, Cohn_Parrish:2021, Choi_Izmaylov:2023} the electronic Hamiltonian in the second-quantized form, $\hat{H}_{e}$, is partitioned into a sum of HF solvable terms, which have eigenstates that are Slater determinants. This partitioning can be achieved by using either low-rank (LR)\cite{Berry_Babbush:2019, Motta_Chan:2021, Huggins_Babbush:2021} or full-rank (FR)\cite{Yen_Izmaylov:2021} decomposition techniques (see Appendix~A of Ref.~\citenum{Choi_Izmaylov:2023} for details). These HF solvable terms have the following form: $\hat{U}_{\alpha} p_{\alpha}(\hat{n}) \hat{U}_{\alpha}^{\dagger}$, where $\hat{n}_{p} = \hat{a}_{p}^{\dagger} \hat{a}_{p}$ is the occupation number operator, and $\hat{U}_{\alpha} = \exp[\sum_{p>q}^{N} \theta_{pq}^{(\alpha)} (\hat{E}^{p}_{q} - \hat{E}^{q}_{p})]$ is the orbital rotation. The $p_{\alpha}(\hat{n})$ polynomial is either a linear function $\sum_{p} \epsilon_{p} \hat{n}_{p}$ (for the one-electron term; $\alpha = 0$) or a quadratic function $\sum_{pq} \lambda_{pq}^{(\alpha)} \hat{n}_{p} \hat{n}_{q}$ (for two-electron terms; $\alpha > 0$) of occupation number operators. Each such fragment is easily measurable since the number operators are mapped onto all-$\hat{z}$ operators under standard fermion-qubit mappings, including Jordan--Wigner (JW) and Bravyi--Kitaev (BK) transformations,\cite{Bravyi_Kitaev:2002, Seeley_Love:2012} and the orbital rotations can be implemented on a quantum circuit with $N(N-1)/2$ Givens rotations and circuit depth linear in $N$.\cite{Kivlichan_Babbush:2018} A recent development\cite{Yordanov_Crispin:2020} discusses an efficient circuit implementation of Givens rotations with a low two-qubit CNOT gate count, thereby improving the practicality of fermionic operator algebra-based methods on near-term quantum devices.

The fluid fermionic fragments (F$^3$) technique has one of the lowest $M(\epsilon)$ in ground state VQE.\cite{Choi_Izmaylov:2023} F$^3$ uses some freedom in the definition of the fragments to
optimize their variances. The main feature used in fragment optimization is the idempotency of the occupation number operators, which allows one to collect a fraction of the two-electron fragments ($c_{\alpha} \sum_{p} \lambda_{pp}^{(\alpha)} \hat{n}_{p}^{2} \equiv c_{\alpha} \sum_{p} \lambda_{pp}^{(\alpha)} \hat{n}_{p}$) into a purely one-electron fragment. The collected one-electron parts can be measured in a single measurement frame defined by a new unitary operator, $\hat{\tilde{U}}$:
\begin{align}
\hat{U}_0^{\dagger} \left( \sum_{p} \epsilon_{p} \hat{n}_{p} \right) \hat{U}_{0} + &\sum_{\alpha} c_{\alpha}  \hat{U}_{\alpha}^{\dagger} \left( \sum_{p} \lambda_{pp}^{(\alpha)} \hat{n}_{p}  \right) \hat{U}_{\alpha} \nonumber \\ &= \hat{\tilde{U}}^{\dagger} \left( \sum_{p} \tilde{\epsilon}_{p} \hat{n}_{p} \right) \hat{\tilde{U}}.
\end{align}
This collection of one-electron parts modifies variances of all fragments, and the $c_{\alpha}$ coefficients are chosen to minimize an approximation to $M(\epsilon)$ obtained using a classically efficient proxy (CISD wavefunction).

Alternatively to the fermionic techniques, the qubit algebra-based techniques exploit that a set of mutually commutative Pauli products can be simultaneously turned into all-$\hat{z}$ operators. This transformation is achieved by a single Clifford rotation,\cite{Yen_Izmaylov:2020, Shlosberg_Jena:2021, Crawford_Brierley:2021, Choi_Izmaylov:2022, Yen_Izmaylov:2023} easily implementable on a quantum computer.\cite{Gottesman:1998, Aaronson_Gottesman:2004,book_Nielsen_Chuang} If one imposes a stricter condition that $\hat{H}_{\alpha}$ are composed only of mutually qubit-wise commuting (QWC)\cite{Verteletskyi_Izmaylov:2020} Pauli products, then the corresponding $\hat{U}_{\alpha}$ diagonalizing $\hat{H}_{\alpha}$ is a tensor product of simpler-to-implement single-qubit Clifford rotations. While the increased freedom in finding the fully commuting (FC) fragments leads to a lower $M(\epsilon)$, the quantum gate error introduced by $\hat{U}_{\alpha}$ is smaller in the QWC scheme due to the absence of low fidelity two-qubit gates. A recent study\cite{Bansingh_Izmaylov:2022} showed that FC techniques typically achieve a smaller number of measurements than that achieved by QWC techniques, even when the non-unit fidelities of quantum gates implementing $\hat{U}_{\alpha}$ are taken into account. 

The classical shadow (CS) methods based on qubit algebra chooses a set of $\hat{U}_{\alpha}$ uniformly at random from all $N$-qubit Clifford rotations (in FC-CS) or a tensor product of single-qubit Clifford rotations (in QWC-CS). The corresponding $\hat{H}_{\alpha}$ is then formed from every Pauli product in $\hat{H}_{q}$ rotated by $\hat{U}_{\alpha}$ into a Pauli-$\hat{z}$ operator.\cite{Huang_Preskill:2020} While such an approach is appropriate for measuring a large number of observables simultaneously, when one desires to measure the expectation value of a single operator, the $M(\epsilon)$ metric [Eq.~(\ref{eq:metric})] can be reduced significantly by choosing $\hat{U}_{\alpha}$'s to prioritize the measurement of $\hat{P}_{j}$'s with large coefficients in $\hat{H}_{q}$. Such reduction in $M(\epsilon)$ was achieved in the derandomized extension of QWC-CS (Derand).\cite{Huang_Preskill:2021} 

The sorted insertion (SI) algorithm\cite{Crawford_Brierley:2021} is a heuristic algorithm that also uses the knowledge of $|c_{j}|$ in $\hat{H}_{q}$ to obtain measurable $\hat{H}_{\alpha}$ fragments with a low $M(\epsilon)$. As a ``greedy'' algorithm, SI naturally leads to a set of fragments with an uneven distribution of variances. The advantage of this uneven distribution is revealed by making the optimal choice of $m_{\alpha} = [\mathrm{Var}_{\psi}(\hat{H}_{\alpha})]^{1/2} / \sum_{\beta} [\mathrm{Var}_{\psi}(\hat{H}_{\beta})]^{1/2}$, which minimizes $M(\epsilon)$ in the ground state VQE. This choice of $m_{\alpha}$ yields
\begin{equation}
M_{\mathrm{opt}}(\epsilon) = \frac{1}{\epsilon^{2}} \left[ \sum_{\beta} \sqrt{\mathrm{Var}_{\psi}(\hat{H}_{\beta})} \right]^{2},
\end{equation} 
and for a fixed sum of variances, the sum of square roots in $M_{\mathrm{opt}}(\epsilon)$ is lower when the variances are distributed unevenly.

While $M(\epsilon)$ in SI is typically lower than those in CS and Derand for ground state VQE, Yen \textit{et al.}\cite{Yen_Izmaylov:2023} lowered it even further by taking advantage of that some $\hat{P}_{j}$'s can be placed in multiple $\hat{H}_{\alpha}$. Two different algorithms for optimizing the amount of $c_{j} \hat{P}_{j}$ in each $\hat{H}_{\alpha}$ were proposed. The iterative coefficient splitting (ICS) has a lower $M(\epsilon)$ but has more parameters to optimize, whereas the iterative measurement allocation (IMA) has a lower classical cost but a higher quantum measurement cost [$M(\epsilon)$].\cite{Yen_Izmaylov:2023}

Another important class of measurement techniques relies on Majorana operator algebra.\cite{Bonet-Monroig_OBrien:2020, Zhao_Miyake:2021} Using the definition of Majorana operators,
\begin{align}
\hat{\gamma}_{2p} &= \hat{a}_{p} + \hat{a}_{p}^{\dagger} \nonumber \\
\hat{\gamma}_{2p + 1} &= - i (\hat{a}_{p} - \hat{a}_{p}^{\dagger}),
\end{align}
for $p = 1, \dots, N$, one can re-express $\hat{H}_{e}$ as a quartic polynomial of $\hat{\gamma}_{p}$. This Majorana Hamiltonian can be partitioned into measurable $\hat{H}_{\alpha}$ fragments by using that under the JW transformation, Majorana products with a $\hat{\gamma}_{2p}\hat{\gamma}_{2p+1}$ form map onto a Pauli-$\hat{z}$ operator. In Majorana operator algebra-based measurement schemes, each $\hat{H}_{\alpha}$ fragment consists of a set of Majorana terms that can be transformed into polynomials of $\hat{\gamma}_{2p}\hat{\gamma}_{2p+1}$ via a unitary operator that permutes the indices of the Majoranas. Such unitary operators can easily be implemented on a quantum computer because they are from a subset of Clifford rotations.\cite{Bonet-Monroig_OBrien:2020} The Majorana-algebra-based CS method,\cite{Zhao_Miyake:2021} which chooses $\hat{U}_{\alpha}$ randomly from a set of Majorana-index permuting operators, was shown to be particularly efficient for simultaneously measuring all elements of a $k$-RDM.

Instead of finding the $\hat{H}_{\alpha}$ fragments that can be easily rotated into measurable all-$\hat{z}$ operators, there also exist positive operator-valued measure (POVM) based techniques. These methods embed $\hat{H}_{q}$ into a larger 2$N$-qubit Hilbert space such that the extended Hamiltonian is easily diagonalizable into the Ising form by a single unitary operator.\cite{Garcia-Perez_Maniscalco:2021, Glos_Garcia-Perez:2022} Note that this scheme only addresses the measurement problem and does not solve the original electronic structure problem because the eigenvalues of the original $\hat{H}_{q}$ are not preserved by the embedding.

\section*{Appendix B: Notation in QSE explained through a simple example}
\label{appendixb}
For simplicity, let us consider a simple case where we have two operators to measure (i.e., $N_{\mathrm{op}} = 2$):
\begin{align}
\hat{A}_{1} &= c_{1,1} \hat{P}_{1} + c_{1,3} \hat{P}_{3}, \nonumber \\
\hat{A}_{2} &= c_{2,1} \hat{P}_{1} + c_{2,2} \hat{P}_{2} + c_{2,3} \hat{P}_{3};
\end{align}
$\hat{P}_{1} = \hat{z}_{1}, \hat{P}_{2} = \hat{z}_{1}\hat{z}_{2}$, and $\hat{P}_{3} = \hat{x}_{1} \hat{x}_{2}$. In this example, $\hat{P}_{2}$ commutes with both $\hat{P}_{1}$ and $\hat{P}_{3}$, but $\hat{P}_{1}$ and $\hat{P}_{3}$ do not commute with each other, and $\mathcal{A}_{1} = \{1, 3\}$ and $\mathcal{A}_{2} = \{1, 2, 3\}$. One possible way to partition $\hat{P}_{k}$'s into simultaneously measurable sets (e.g., in FC-SI) is to measure $\hat{P}_{1}$ and $\hat{P}_{2}$ together and measure $\hat{P}_{3}$ separately. Under this partitioning, $\mathcal{S}_{1} = \{1, 2\}$ and $\mathcal{S}_{2} = \{3\}$. Then, the measurable fragments for $\hat{A}_{1}$ are
\begin{equation}
\hat{A}_{1}^{(1)} = c_{1,1} \hat{P}_{1}, \quad 
\hat{A}_{1}^{(2)} = c_{1,3} \hat{P}_{3},
\end{equation}
with $\mathcal{A}_{1} \cap \mathcal{S}_{1} = \{1\}$ and $\mathcal{A}_{1} \cap \mathcal{S}_{2} = \{3\}$. For $\hat{A}_{2}$, the fragments are
\begin{equation}
\hat{A}_{2}^{(1)} = c_{2,1} \hat{P}_{1} + c_{2,2} \hat{P}_{2}, \quad
\hat{A}_{2}^{(2)} = c_{2,3} \hat{P}_{3},
\end{equation}
with $\mathcal{A}_{2} \cap \mathcal{S}_{1} = \{1, 2\}$ and $\mathcal{A}_{2} \cap \mathcal{S}_{2} = \{3\}$.

\bibliography{ExcitedStates}

\providecommand{\latin}[1]{#1}
\makeatletter
\providecommand{\doi}
  {\begingroup\let\do\@makeother\dospecials
  \catcode`\{=1 \catcode`\}=2 \doi@aux}
\providecommand{\doi@aux}[1]{\endgroup\texttt{#1}}
\makeatother
\providecommand*\mcitethebibliography{\thebibliography}
\csname @ifundefined\endcsname{endmcitethebibliography}
  {\let\endmcitethebibliography\endthebibliography}{}
\begin{mcitethebibliography}{64}
\providecommand*\natexlab[1]{#1}
\providecommand*\mciteSetBstSublistMode[1]{}
\providecommand*\mciteSetBstMaxWidthForm[2]{}
\providecommand*\mciteBstWouldAddEndPuncttrue
  {\def\EndOfBibitem{\unskip.}}
\providecommand*\mciteBstWouldAddEndPunctfalse
  {\let\EndOfBibitem\relax}
\providecommand*\mciteSetBstMidEndSepPunct[3]{}
\providecommand*\mciteSetBstSublistLabelBeginEnd[3]{}
\providecommand*\EndOfBibitem{}
\mciteSetBstSublistMode{f}
\mciteSetBstMaxWidthForm{subitem}{(\alph{mcitesubitemcount})}
\mciteSetBstSublistLabelBeginEnd
  {\mcitemaxwidthsubitemform\space}
  {\relax}
  {\relax}

\bibitem[Peruzzo \latin{et~al.}(2014)Peruzzo, McClean, Shadbolt, Yung, Zhou,
  Love, Aspuru-Guzik, and O’Brien]{Peruzzo_OBrien:2014}
Peruzzo,~A.; McClean,~J.; Shadbolt,~P.; Yung,~M.-H.; Zhou,~X.-Q.; Love,~P.~J.;
  Aspuru-Guzik,~A.; O’Brien,~J.~L. A variational eigenvalue solver on a
  photonic quantum processor. \emph{Nat.~Commun.} \textbf{2014}, \emph{5},
  1--7\relax
\mciteBstWouldAddEndPuncttrue
\mciteSetBstMidEndSepPunct{\mcitedefaultmidpunct}
{\mcitedefaultendpunct}{\mcitedefaultseppunct}\relax
\EndOfBibitem
\bibitem[McClean \latin{et~al.}(2016)McClean, Romero, Babbush, and
  Aspuru-Guzik]{McClean_Aspuru-Guzik:2016}
McClean,~J.~R.; Romero,~J.; Babbush,~R.; Aspuru-Guzik,~A. The theory of
  variational hybrid quantum-classical algorithms. \emph{New J.~Phys.}
  \textbf{2016}, \emph{18}, 023023\relax
\mciteBstWouldAddEndPuncttrue
\mciteSetBstMidEndSepPunct{\mcitedefaultmidpunct}
{\mcitedefaultendpunct}{\mcitedefaultseppunct}\relax
\EndOfBibitem
\bibitem[Ryabinkin \latin{et~al.}(2020)Ryabinkin, Lang, Genin, and
  Izmaylov]{Rybinkin_Izmaylov:2020}
Ryabinkin,~I.~G.; Lang,~R.~A.; Genin,~S.~N.; Izmaylov,~A.~F. Iterative Qubit
  Coupled Cluster Approach with Efficient Screening of Generators.
  \emph{J.~Chem.\ Theory Comput.} \textbf{2020}, \emph{16}, 1055--1063\relax
\mciteBstWouldAddEndPuncttrue
\mciteSetBstMidEndSepPunct{\mcitedefaultmidpunct}
{\mcitedefaultendpunct}{\mcitedefaultseppunct}\relax
\EndOfBibitem
\bibitem[Cerezo \latin{et~al.}(2021)Cerezo, Arrasmith, Babbush, Benjamin, Endo,
  Fujii, McClean, Mitarai, Yuan, Cincio, and Coles]{Cerezo_Coles:2021}
Cerezo,~M.; Arrasmith,~A.; Babbush,~R.; Benjamin,~S.~C.; Endo,~S.; Fujii,~K.;
  McClean,~J.~R.; Mitarai,~K.; Yuan,~X.; Cincio,~L.; Coles,~P.~J. Variational
  quantum algorithms. \emph{Nat.~Rev.~Phys.} \textbf{2021}, \emph{3},
  625--644\relax
\mciteBstWouldAddEndPuncttrue
\mciteSetBstMidEndSepPunct{\mcitedefaultmidpunct}
{\mcitedefaultendpunct}{\mcitedefaultseppunct}\relax
\EndOfBibitem
\bibitem[Anand \latin{et~al.}(2022)Anand, Schleich, Alperin-Lea, Jensen, Sim,
  D{\'i}az-Tinoco, Kottmann, Degroote, Izmaylov, and
  Aspuru-Guzik]{Anand_Aspuru-Guzik:2022}
Anand,~A.; Schleich,~P.; Alperin-Lea,~S.; Jensen,~P. W.~K.; Sim,~S.;
  D{\'i}az-Tinoco,~M.; Kottmann,~J.~S.; Degroote,~M.; Izmaylov,~A.~F.;
  Aspuru-Guzik,~A. A quantum computing view on unitary coupled cluster theory.
  \emph{Chem. Soc. Rev.} \textbf{2022}, \emph{51}, 1659--1684\relax
\mciteBstWouldAddEndPuncttrue
\mciteSetBstMidEndSepPunct{\mcitedefaultmidpunct}
{\mcitedefaultendpunct}{\mcitedefaultseppunct}\relax
\EndOfBibitem
\bibitem[Preskill(2018)]{Preskill:2018}
Preskill,~J. Quantum Computing in the {NISQ} era and beyond. \emph{{Quantum}}
  \textbf{2018}, \emph{2}, 79\relax
\mciteBstWouldAddEndPuncttrue
\mciteSetBstMidEndSepPunct{\mcitedefaultmidpunct}
{\mcitedefaultendpunct}{\mcitedefaultseppunct}\relax
\EndOfBibitem
\bibitem[Gonthier \latin{et~al.}(2022)Gonthier, Radin, Buda, Doskocil, Abuan,
  and Romero]{Gonthier_Romero:2022}
Gonthier,~J.~F.; Radin,~M.~D.; Buda,~C.; Doskocil,~E.~J.; Abuan,~C.~M.;
  Romero,~J. Measurements as a roadblock to near-term practical quantum
  advantage in chemistry: Resource analysis. \emph{Phys. Rev. Research}
  \textbf{2022}, \emph{4}, 033154\relax
\mciteBstWouldAddEndPuncttrue
\mciteSetBstMidEndSepPunct{\mcitedefaultmidpunct}
{\mcitedefaultendpunct}{\mcitedefaultseppunct}\relax
\EndOfBibitem
\bibitem[Crawford \latin{et~al.}(2021)Crawford, Straaten, Wang, Parks,
  Campbell, and Brierley]{Crawford_Brierley:2021}
Crawford,~O.; Straaten,~B.~v.; Wang,~D.; Parks,~T.; Campbell,~E.; Brierley,~S.
  Efficient quantum measurement of {P}auli operators in the presence of finite
  sampling error. \emph{{Quantum}} \textbf{2021}, \emph{5}, 385\relax
\mciteBstWouldAddEndPuncttrue
\mciteSetBstMidEndSepPunct{\mcitedefaultmidpunct}
{\mcitedefaultendpunct}{\mcitedefaultseppunct}\relax
\EndOfBibitem
\bibitem[Yen \latin{et~al.}(2023)Yen, Ganeshram, and
  Izmaylov]{Yen_Izmaylov:2023}
Yen,~T.-C.; Ganeshram,~A.; Izmaylov,~A.~F. Deterministic improvements of
  quantum measurements with grouping of compatible operators, non-local
  transformations, and covariance estimates. \emph{npj Quantum Inf.}
  \textbf{2023}, \emph{9}, 14\relax
\mciteBstWouldAddEndPuncttrue
\mciteSetBstMidEndSepPunct{\mcitedefaultmidpunct}
{\mcitedefaultendpunct}{\mcitedefaultseppunct}\relax
\EndOfBibitem
\bibitem[Huggins \latin{et~al.}(2021)Huggins, McClean, Rubin, Jiang, Wiebe,
  Whaley, and Babbush]{Huggins_Babbush:2021}
Huggins,~W.~J.; McClean,~J.~R.; Rubin,~N.~C.; Jiang,~Z.; Wiebe,~N.;
  Whaley,~K.~B.; Babbush,~R. Efficient and noise resilient measurements for
  quantum chemistry on near-term quantum computers. \emph{npj Quantum Inf.}
  \textbf{2021}, \emph{7}, 1--9\relax
\mciteBstWouldAddEndPuncttrue
\mciteSetBstMidEndSepPunct{\mcitedefaultmidpunct}
{\mcitedefaultendpunct}{\mcitedefaultseppunct}\relax
\EndOfBibitem
\bibitem[Yen and Izmaylov(2021)Yen, and Izmaylov]{Yen_Izmaylov:2021}
Yen,~T.-C.; Izmaylov,~A.~F. Cartan Subalgebra Approach to Efficient
  Measurements of Quantum Observables. \emph{PRX Quantum} \textbf{2021},
  \emph{2}, 040320\relax
\mciteBstWouldAddEndPuncttrue
\mciteSetBstMidEndSepPunct{\mcitedefaultmidpunct}
{\mcitedefaultendpunct}{\mcitedefaultseppunct}\relax
\EndOfBibitem
\bibitem[Cohn \latin{et~al.}(2021)Cohn, Motta, and Parrish]{Cohn_Parrish:2021}
Cohn,~J.; Motta,~M.; Parrish,~R.~M. Quantum Filter Diagonalization with
  Compressed Double-Factorized Hamiltonians. \emph{PRX Quantum} \textbf{2021},
  \emph{2}, 040352\relax
\mciteBstWouldAddEndPuncttrue
\mciteSetBstMidEndSepPunct{\mcitedefaultmidpunct}
{\mcitedefaultendpunct}{\mcitedefaultseppunct}\relax
\EndOfBibitem
\bibitem[Choi \latin{et~al.}(2023)Choi, Loaiza, and
  Izmaylov]{Choi_Izmaylov:2023}
Choi,~S.; Loaiza,~I.; Izmaylov,~A.~F. Fluid fermionic fragments for optimizing
  quantum measurements of electronic {H}amiltonians in the variational quantum
  eigensolver. \emph{{Quantum}} \textbf{2023}, \emph{7}, 889\relax
\mciteBstWouldAddEndPuncttrue
\mciteSetBstMidEndSepPunct{\mcitedefaultmidpunct}
{\mcitedefaultendpunct}{\mcitedefaultseppunct}\relax
\EndOfBibitem
\bibitem[Yen \latin{et~al.}(2020)Yen, Verteletskyi, and
  Izmaylov]{Yen_Izmaylov:2020}
Yen,~T.-C.; Verteletskyi,~V.; Izmaylov,~A.~F. Measuring All Compatible
  Operators in One Series of Single-Qubit Measurements Using Unitary
  Transformations. \emph{J.~Chem.\ Theory Comput.} \textbf{2020}, \emph{16},
  2400--2409\relax
\mciteBstWouldAddEndPuncttrue
\mciteSetBstMidEndSepPunct{\mcitedefaultmidpunct}
{\mcitedefaultendpunct}{\mcitedefaultseppunct}\relax
\EndOfBibitem
\bibitem[Verteletskyi \latin{et~al.}(2020)Verteletskyi, Yen, and
  Izmaylov]{Verteletskyi_Izmaylov:2020}
Verteletskyi,~V.; Yen,~T.-C.; Izmaylov,~A.~F. Measurement optimization in the
  variational quantum eigensolver using a minimum clique cover. \emph{J.~Chem.\
  Phys.} \textbf{2020}, \emph{152}, 124114\relax
\mciteBstWouldAddEndPuncttrue
\mciteSetBstMidEndSepPunct{\mcitedefaultmidpunct}
{\mcitedefaultendpunct}{\mcitedefaultseppunct}\relax
\EndOfBibitem
\bibitem[Choi \latin{et~al.}(2022)Choi, Yen, and Izmaylov]{Choi_Izmaylov:2022}
Choi,~S.; Yen,~T.-C.; Izmaylov,~A.~F. Improving Quantum Measurements by
  Introducing “Ghost” Pauli Products. \emph{J.~Chem.\ Theory Comput.}
  \textbf{2022}, \emph{18}, 7394--7402\relax
\mciteBstWouldAddEndPuncttrue
\mciteSetBstMidEndSepPunct{\mcitedefaultmidpunct}
{\mcitedefaultendpunct}{\mcitedefaultseppunct}\relax
\EndOfBibitem
\bibitem[Bonet-Monroig \latin{et~al.}(2020)Bonet-Monroig, Babbush, and
  O'Brien]{Bonet-Monroig_OBrien:2020}
Bonet-Monroig,~X.; Babbush,~R.; O'Brien,~T.~E. Nearly Optimal Measurement
  Scheduling for Partial Tomography of Quantum States. \emph{Phys. Rev. X}
  \textbf{2020}, \emph{10}, 031064\relax
\mciteBstWouldAddEndPuncttrue
\mciteSetBstMidEndSepPunct{\mcitedefaultmidpunct}
{\mcitedefaultendpunct}{\mcitedefaultseppunct}\relax
\EndOfBibitem
\bibitem[Gresch and Kliesch(2023)Gresch, and Kliesch]{Gresch_Kliesch:2023}
Gresch,~A.; Kliesch,~M. Guaranteed efficient energy estimation of quantum
  many-body {H}amiltonians using {ShadowGrouping}. \emph{arXiv:2301.03385}
  \textbf{2023}, \relax
\mciteBstWouldAddEndPunctfalse
\mciteSetBstMidEndSepPunct{\mcitedefaultmidpunct}
{}{\mcitedefaultseppunct}\relax
\EndOfBibitem
\bibitem[Hadfield \latin{et~al.}(2022)Hadfield, Bravyi, Raymond, and
  Mezzacapo]{Hadfield_Mezzacapo:2022}
Hadfield,~C.; Bravyi,~S.; Raymond,~R.; Mezzacapo,~A. Measurements of quantum
  hamiltonians with locally-biased classical shadows. \emph{Commun.\ Math.\
  Phys.} \textbf{2022}, \emph{391}, 951--967\relax
\mciteBstWouldAddEndPuncttrue
\mciteSetBstMidEndSepPunct{\mcitedefaultmidpunct}
{\mcitedefaultendpunct}{\mcitedefaultseppunct}\relax
\EndOfBibitem
\bibitem[Huang \latin{et~al.}(2020)Huang, Kueng, and
  Preskill]{Huang_Preskill:2020}
Huang,~H.-Y.; Kueng,~R.; Preskill,~J. Predicting many properties of a quantum
  system from very few measurements. \emph{Nat.~Phys.} \textbf{2020},
  \emph{16}, 1050--1057\relax
\mciteBstWouldAddEndPuncttrue
\mciteSetBstMidEndSepPunct{\mcitedefaultmidpunct}
{\mcitedefaultendpunct}{\mcitedefaultseppunct}\relax
\EndOfBibitem
\bibitem[Huang \latin{et~al.}(2021)Huang, Kueng, and
  Preskill]{Huang_Preskill:2021}
Huang,~H.-Y.; Kueng,~R.; Preskill,~J. Efficient Estimation of {P}auli
  Observables by Derandomization. \emph{Phys.~Rev.~Lett.} \textbf{2021},
  \emph{127}, 030503\relax
\mciteBstWouldAddEndPuncttrue
\mciteSetBstMidEndSepPunct{\mcitedefaultmidpunct}
{\mcitedefaultendpunct}{\mcitedefaultseppunct}\relax
\EndOfBibitem
\bibitem[Hillmich \latin{et~al.}(2021)Hillmich, Hadfield, Raymond, Mezzacapo,
  and Wille]{Hilmich_Wille:2021}
Hillmich,~S.; Hadfield,~C.; Raymond,~R.; Mezzacapo,~A.; Wille,~R. Decision
  Diagrams for Quantum Measurements with Shallow Circuits. 2021 IEEE
  International Conference on Quantum Computing and Engineering (QCE). 2021; pp
  24--34\relax
\mciteBstWouldAddEndPuncttrue
\mciteSetBstMidEndSepPunct{\mcitedefaultmidpunct}
{\mcitedefaultendpunct}{\mcitedefaultseppunct}\relax
\EndOfBibitem
\bibitem[Wu \latin{et~al.}(2021)Wu, Sun, Huang, and Yuan]{Wu_Yuan:2021}
Wu,~B.; Sun,~J.; Huang,~Q.; Yuan,~X. Overlapped grouping measurement: A unified
  framework for measuring quantum states. \emph{arXiv:2105.13091}
  \textbf{2021}, \relax
\mciteBstWouldAddEndPunctfalse
\mciteSetBstMidEndSepPunct{\mcitedefaultmidpunct}
{}{\mcitedefaultseppunct}\relax
\EndOfBibitem
\bibitem[Hadfield(2021)]{Hadfield:2021}
Hadfield,~C. Adaptive {P}auli Shadows for Energy Estimation.
  \emph{arXiv:2105.12207} \textbf{2021}, \relax
\mciteBstWouldAddEndPunctfalse
\mciteSetBstMidEndSepPunct{\mcitedefaultmidpunct}
{}{\mcitedefaultseppunct}\relax
\EndOfBibitem
\bibitem[Zhao \latin{et~al.}(2021)Zhao, Rubin, and Miyake]{Zhao_Miyake:2021}
Zhao,~A.; Rubin,~N.~C.; Miyake,~A. Fermionic Partial Tomography via Classical
  Shadows. \emph{Phys. Rev. Lett.} \textbf{2021}, \emph{127}, 110504\relax
\mciteBstWouldAddEndPuncttrue
\mciteSetBstMidEndSepPunct{\mcitedefaultmidpunct}
{\mcitedefaultendpunct}{\mcitedefaultseppunct}\relax
\EndOfBibitem
\bibitem[McClean \latin{et~al.}(2017)McClean, Kimchi-Schwartz, Carter, and
  de~Jong]{McClean_deJong:2017}
McClean,~J.~R.; Kimchi-Schwartz,~M.~E.; Carter,~J.; de~Jong,~W.~A. Hybrid
  quantum-classical hierarchy for mitigation of decoherence and determination
  of excited states. \emph{Phys. Rev. A} \textbf{2017}, \emph{95}, 042308\relax
\mciteBstWouldAddEndPuncttrue
\mciteSetBstMidEndSepPunct{\mcitedefaultmidpunct}
{\mcitedefaultendpunct}{\mcitedefaultseppunct}\relax
\EndOfBibitem
\bibitem[Colless \latin{et~al.}(2018)Colless, Ramasesh, Dahlen, Blok,
  Kimchi-Schwartz, McClean, Carter, de~Jong, and Siddiqi]{Colless_Siddiqi:2018}
Colless,~J.~I.; Ramasesh,~V.~V.; Dahlen,~D.; Blok,~M.~S.;
  Kimchi-Schwartz,~M.~E.; McClean,~J.~R.; Carter,~J.; de~Jong,~W.~A.;
  Siddiqi,~I. Computation of Molecular Spectra on a Quantum Processor with an
  Error-Resilient Algorithm. \emph{Phys. Rev. X} \textbf{2018}, \emph{8},
  011021\relax
\mciteBstWouldAddEndPuncttrue
\mciteSetBstMidEndSepPunct{\mcitedefaultmidpunct}
{\mcitedefaultendpunct}{\mcitedefaultseppunct}\relax
\EndOfBibitem
\bibitem[Takeshita \latin{et~al.}(2020)Takeshita, Rubin, Jiang, Lee, Babbush,
  and McClean]{Takeshita_McClean:2020}
Takeshita,~T.; Rubin,~N.~C.; Jiang,~Z.; Lee,~E.; Babbush,~R.; McClean,~J.~R.
  Increasing the Representation Accuracy of Quantum Simulations of Chemistry
  without Extra Quantum Resources. \emph{Phys. Rev. X} \textbf{2020},
  \emph{10}, 011004\relax
\mciteBstWouldAddEndPuncttrue
\mciteSetBstMidEndSepPunct{\mcitedefaultmidpunct}
{\mcitedefaultendpunct}{\mcitedefaultseppunct}\relax
\EndOfBibitem
\bibitem[McClean \latin{et~al.}(2020)McClean, Jiang, Rubin, Babbush, and
  Neven]{McClean_Neven:2020}
McClean,~J.~R.; Jiang,~Z.; Rubin,~N.~C.; Babbush,~R.; Neven,~H. Decoding
  quantum errors with subspace expansions. \emph{Nat.~Commun.} \textbf{2020},
  \emph{11}, 1--9\relax
\mciteBstWouldAddEndPuncttrue
\mciteSetBstMidEndSepPunct{\mcitedefaultmidpunct}
{\mcitedefaultendpunct}{\mcitedefaultseppunct}\relax
\EndOfBibitem
\bibitem[Yoshioka \latin{et~al.}(2022)Yoshioka, Hakoshima, Matsuzaki, Tokunaga,
  Suzuki, and Endo]{Yoshioka_Endo:2022}
Yoshioka,~N.; Hakoshima,~H.; Matsuzaki,~Y.; Tokunaga,~Y.; Suzuki,~Y.; Endo,~S.
  Generalized Quantum Subspace Expansion. \emph{Phys. Rev. Lett.}
  \textbf{2022}, \emph{129}, 020502\relax
\mciteBstWouldAddEndPuncttrue
\mciteSetBstMidEndSepPunct{\mcitedefaultmidpunct}
{\mcitedefaultendpunct}{\mcitedefaultseppunct}\relax
\EndOfBibitem
\bibitem[Parrish \latin{et~al.}(2019)Parrish, Hohenstein, McMahon, and
  Mart\'{\i}nez]{Parrish_Martinez:2019}
Parrish,~R.~M.; Hohenstein,~E.~G.; McMahon,~P.~L.; Mart\'{\i}nez,~T.~J. Quantum
  Computation of Electronic Transitions Using a Variational Quantum
  Eigensolver. \emph{Phys. Rev. Lett.} \textbf{2019}, \emph{122}, 230401\relax
\mciteBstWouldAddEndPuncttrue
\mciteSetBstMidEndSepPunct{\mcitedefaultmidpunct}
{\mcitedefaultendpunct}{\mcitedefaultseppunct}\relax
\EndOfBibitem
\bibitem[Parrish \latin{et~al.}(2019)Parrish, Hohenstein, McMahon, and
  Martinez]{Parrish_Martinez:2019a}
Parrish,~R.~M.; Hohenstein,~E.~G.; McMahon,~P.~L.; Martinez,~T.~J. Hybrid
  Quantum/Classical Derivative Theory: Analytical Gradients and Excited-State
  Dynamics for the Multistate Contracted Variational Quantum Eigensolver.
  \emph{arXiv:1906.08728} \textbf{2019}, \relax
\mciteBstWouldAddEndPunctfalse
\mciteSetBstMidEndSepPunct{\mcitedefaultmidpunct}
{}{\mcitedefaultseppunct}\relax
\EndOfBibitem
\bibitem[Urbanek \latin{et~al.}(2020)Urbanek, Camps, Van~Beeumen, and
  de~Jong]{Urbaneck_deJong:2020}
Urbanek,~M.; Camps,~D.; Van~Beeumen,~R.; de~Jong,~W.~A. Chemistry on Quantum
  Computers with Virtual Quantum Subspace Expansion. \emph{J.~Chem.\ Theory
  Comput.} \textbf{2020}, \emph{16}, 5425--5431\relax
\mciteBstWouldAddEndPuncttrue
\mciteSetBstMidEndSepPunct{\mcitedefaultmidpunct}
{\mcitedefaultendpunct}{\mcitedefaultseppunct}\relax
\EndOfBibitem
\bibitem[Huang \latin{et~al.}(2022)Huang, Govoni, and Galli]{Huang_Galli:2022}
Huang,~B.; Govoni,~M.; Galli,~G. Simulating the Electronic Structure of Spin
  Defects on Quantum Computers. \emph{PRX Quantum} \textbf{2022}, \emph{3},
  010339\relax
\mciteBstWouldAddEndPuncttrue
\mciteSetBstMidEndSepPunct{\mcitedefaultmidpunct}
{\mcitedefaultendpunct}{\mcitedefaultseppunct}\relax
\EndOfBibitem
\bibitem[Tammaro \latin{et~al.}(2022)Tammaro, Galli, Rice, and
  Motta]{Tammaro_Motta:2022}
Tammaro,~A.; Galli,~D.~E.; Rice,~J.~E.; Motta,~M. N-electron valence
  perturbation theory with reference wavefunctions from quantum computing:
  application to the relative stability of hydroxide anion and hydroxyl
  radical. \emph{arXiv:2202.13002} \textbf{2022}, \relax
\mciteBstWouldAddEndPunctfalse
\mciteSetBstMidEndSepPunct{\mcitedefaultmidpunct}
{}{\mcitedefaultseppunct}\relax
\EndOfBibitem
\bibitem[Barison \latin{et~al.}(2022)Barison, Galli, and
  Motta]{Barison_Motta:2022}
Barison,~S.; Galli,~D.~E.; Motta,~M. Quantum simulations of molecular systems
  with intrinsic atomic orbitals. \emph{Phys. Rev. A} \textbf{2022},
  \emph{106}, 022404\relax
\mciteBstWouldAddEndPuncttrue
\mciteSetBstMidEndSepPunct{\mcitedefaultmidpunct}
{\mcitedefaultendpunct}{\mcitedefaultseppunct}\relax
\EndOfBibitem
\bibitem[L\"otstedt \latin{et~al.}(2021)L\"otstedt, Yamanouchi, Tsuchiya, and
  Tachikawa]{Lotstedt_Tachikawa:2021}
L\"otstedt,~E.; Yamanouchi,~K.; Tsuchiya,~T.; Tachikawa,~Y. Calculation of
  vibrational eigenenergies on a quantum computer: Application to the Fermi
  resonance in ${\mathrm{CO}}_{2}$. \emph{Phys. Rev. A} \textbf{2021},
  \emph{103}, 062609\relax
\mciteBstWouldAddEndPuncttrue
\mciteSetBstMidEndSepPunct{\mcitedefaultmidpunct}
{\mcitedefaultendpunct}{\mcitedefaultseppunct}\relax
\EndOfBibitem
\bibitem[L{\"o}tstedt \latin{et~al.}(2022)L{\"o}tstedt, Yamanouchi, and
  Tachikawa]{Lotstedt_Tachikawa:2022}
L{\"o}tstedt,~E.; Yamanouchi,~K.; Tachikawa,~Y. Evaluation of vibrational
  energies and wave functions of {CO$_{2}$} on a quantum computer. \emph{AVS
  Quantum Sci.} \textbf{2022}, \emph{4}, 036801\relax
\mciteBstWouldAddEndPuncttrue
\mciteSetBstMidEndSepPunct{\mcitedefaultmidpunct}
{\mcitedefaultendpunct}{\mcitedefaultseppunct}\relax
\EndOfBibitem
\bibitem[Parrish and McMahon(2019)Parrish, and McMahon]{Parrish_McMahon:2019}
Parrish,~R.~M.; McMahon,~P.~L. Quantum Filter Diagonalization: Quantum
  Eigendecomposition without Full Quantum Phase Estimation.
  \emph{arXiv:1909.08925} \textbf{2019}, \relax
\mciteBstWouldAddEndPunctfalse
\mciteSetBstMidEndSepPunct{\mcitedefaultmidpunct}
{}{\mcitedefaultseppunct}\relax
\EndOfBibitem
\bibitem[Huggins \latin{et~al.}(2020)Huggins, Lee, Baek, O’Gorman, and
  Whaley]{Huggins_Whaley:2020}
Huggins,~W.~J.; Lee,~J.; Baek,~U.; O’Gorman,~B.; Whaley,~K.~B. A
  non-orthogonal variational quantum eigensolver. \emph{New J.~Phys.}
  \textbf{2020}, \emph{22}, 073009\relax
\mciteBstWouldAddEndPuncttrue
\mciteSetBstMidEndSepPunct{\mcitedefaultmidpunct}
{\mcitedefaultendpunct}{\mcitedefaultseppunct}\relax
\EndOfBibitem
\bibitem[Motta \latin{et~al.}(2020)Motta, Sun, Tan, O’Rourke, Ye, Minnich,
  Brand{\~a}o, and Chan]{Motta_Chan:2020}
Motta,~M.; Sun,~C.; Tan,~A.~T.; O’Rourke,~M.~J.; Ye,~E.; Minnich,~A.~J.;
  Brand{\~a}o,~F.~G.; Chan,~G. K.-L. Determining eigenstates and thermal states
  on a quantum computer using quantum imaginary time evolution. \emph{Nature
  Physics} \textbf{2020}, \emph{16}, 205--210\relax
\mciteBstWouldAddEndPuncttrue
\mciteSetBstMidEndSepPunct{\mcitedefaultmidpunct}
{\mcitedefaultendpunct}{\mcitedefaultseppunct}\relax
\EndOfBibitem
\bibitem[Stair \latin{et~al.}(2020)Stair, Huang, and
  Evangelista]{Stair_Evangelista:2020}
Stair,~N.~H.; Huang,~R.; Evangelista,~F.~A. A multireference quantum Krylov
  algorithm for strongly correlated electrons. \emph{J.~Chem.\ Theory Comput.}
  \textbf{2020}, \emph{16}, 2236--2245\relax
\mciteBstWouldAddEndPuncttrue
\mciteSetBstMidEndSepPunct{\mcitedefaultmidpunct}
{\mcitedefaultendpunct}{\mcitedefaultseppunct}\relax
\EndOfBibitem
\bibitem[Cortes and Gray(2022)Cortes, and Gray]{Cortes_Gray:2022}
Cortes,~C.~L.; Gray,~S.~K. Quantum Krylov subspace algorithms for ground- and
  excited-state energy estimation. \emph{Phys. Rev. A} \textbf{2022},
  \emph{105}, 022417\relax
\mciteBstWouldAddEndPuncttrue
\mciteSetBstMidEndSepPunct{\mcitedefaultmidpunct}
{\mcitedefaultendpunct}{\mcitedefaultseppunct}\relax
\EndOfBibitem
\bibitem[Chiew and Kwek()Chiew, and Kwek]{Chiew_Kwek:2023}
Chiew,~S.-H.; Kwek,~L.-C. Scalable Quantum Computation of Highly Excited
  Eigenstates with Spectral Transforms. \emph{arXiv:2302.06638} \relax
\mciteBstWouldAddEndPunctfalse
\mciteSetBstMidEndSepPunct{\mcitedefaultmidpunct}
{}{\mcitedefaultseppunct}\relax
\EndOfBibitem
\bibitem[Epperly \latin{et~al.}(2022)Epperly, Lin, and
  Nakatsukasa]{Epperly_Nakatsukasa:2022}
Epperly,~E.~N.; Lin,~L.; Nakatsukasa,~Y. A Theory of Quantum Subspace
  Diagonalization. \emph{SIAM J. Matrix Anal. Appl.} \textbf{2022}, \emph{43},
  1263--1290\relax
\mciteBstWouldAddEndPuncttrue
\mciteSetBstMidEndSepPunct{\mcitedefaultmidpunct}
{\mcitedefaultendpunct}{\mcitedefaultseppunct}\relax
\EndOfBibitem
\bibitem[Mathias and Li(2004)Mathias, and Li]{Mathias_Li:2004}
Mathias,~R.; Li,~C.-K. The definite generalized eigenvalue problem: A new
  perturbation theory. T-NAREP No. 457, inst-MCCM, 2004;
  \url{https://www.maths.manchester.ac.uk/~higham/narep/narep457.pdf}\relax
\mciteBstWouldAddEndPuncttrue
\mciteSetBstMidEndSepPunct{\mcitedefaultmidpunct}
{\mcitedefaultendpunct}{\mcitedefaultseppunct}\relax
\EndOfBibitem
\bibitem[Vershynin(2018)]{Vershynin:2018}
Vershynin,~R. \emph{High-dimensional probability: An introduction with
  applications in data science}; Cambridge university press, 2018; Vol.~47;
  p~85\relax
\mciteBstWouldAddEndPuncttrue
\mciteSetBstMidEndSepPunct{\mcitedefaultmidpunct}
{\mcitedefaultendpunct}{\mcitedefaultseppunct}\relax
\EndOfBibitem
\bibitem[Berry \latin{et~al.}(2019)Berry, Gidney, Motta, McClean, and
  Babbush]{Berry_Babbush:2019}
Berry,~D.~W.; Gidney,~C.; Motta,~M.; McClean,~J.~R.; Babbush,~R. Qubitization
  of arbitrary basis quantum chemistry leveraging sparsity and low rank
  factorization. \emph{Quantum} \textbf{2019}, \emph{3}, 208\relax
\mciteBstWouldAddEndPuncttrue
\mciteSetBstMidEndSepPunct{\mcitedefaultmidpunct}
{\mcitedefaultendpunct}{\mcitedefaultseppunct}\relax
\EndOfBibitem
\bibitem[Motta \latin{et~al.}(2021)Motta, Ye, McClean, Li, Minnich, Babbush,
  and Chan]{Motta_Chan:2021}
Motta,~M.; Ye,~E.; McClean,~J.~R.; Li,~Z.; Minnich,~A.~J.; Babbush,~R.;
  Chan,~G. K.-L. Low rank representations for quantum simulation of electronic
  structure. \emph{npj Quantum Inf.} \textbf{2021}, \emph{7}, 1--7\relax
\mciteBstWouldAddEndPuncttrue
\mciteSetBstMidEndSepPunct{\mcitedefaultmidpunct}
{\mcitedefaultendpunct}{\mcitedefaultseppunct}\relax
\EndOfBibitem
\bibitem[Garc\'{\i}a-P\'erez \latin{et~al.}(2021)Garc\'{\i}a-P\'erez, Rossi,
  Sokolov, Tacchino, Barkoutsos, Mazzola, Tavernelli, and
  Maniscalco]{Garcia-Perez_Maniscalco:2021}
Garc\'{\i}a-P\'erez,~G.; Rossi,~M.~A.; Sokolov,~B.; Tacchino,~F.;
  Barkoutsos,~P.~K.; Mazzola,~G.; Tavernelli,~I.; Maniscalco,~S. Learning to
  Measure: Adaptive Informationally Complete Generalized Measurements for
  Quantum Algorithms. \emph{PRX Quantum} \textbf{2021}, \emph{2}, 040342\relax
\mciteBstWouldAddEndPuncttrue
\mciteSetBstMidEndSepPunct{\mcitedefaultmidpunct}
{\mcitedefaultendpunct}{\mcitedefaultseppunct}\relax
\EndOfBibitem
\bibitem[Glos \latin{et~al.}(2022)Glos, Nyk{\"a}nen, Borrelli, Maniscalco,
  Rossi, Zimbor{\'a}s, and Garc{\'\i}a-P{\'e}rez]{Glos_Garcia-Perez:2022}
Glos,~A.; Nyk{\"a}nen,~A.; Borrelli,~E.-M.; Maniscalco,~S.; Rossi,~M. A.~C.;
  Zimbor{\'a}s,~Z.; Garc{\'\i}a-P{\'e}rez,~G. Adaptive {POVM} implementations
  and measurement error mitigation strategies for near-term quantum devices.
  \emph{arXiv:2208.07817} \textbf{2022}, \relax
\mciteBstWouldAddEndPunctfalse
\mciteSetBstMidEndSepPunct{\mcitedefaultmidpunct}
{}{\mcitedefaultseppunct}\relax
\EndOfBibitem
\bibitem[Choi \latin{et~al.}(2022)Choi, Loaiza, and
  Izmaylov]{Choi_Izmaylov_data:2022}
Choi,~S.; Loaiza,~I.; Izmaylov,~A.~F. {Data for: Fluid fermionic fragments for
  optimizing quantum measurements of electronic Hamiltonians in the variational
  quantum eigensolver}. 2022;
  \url{https://doi.org/10.5281/zenodo.7335451}\relax
\mciteBstWouldAddEndPuncttrue
\mciteSetBstMidEndSepPunct{\mcitedefaultmidpunct}
{\mcitedefaultendpunct}{\mcitedefaultseppunct}\relax
\EndOfBibitem
\bibitem[Sherrill(1996)]{Sherrill:1996}
Sherrill,~C.~D. Computational Scaling of the Configuration Interaction Method
  with System Size. 1996;
  \url{http://vergil.chemistry.gatech.edu/notes/ciscale/ciscale.html}\relax
\mciteBstWouldAddEndPuncttrue
\mciteSetBstMidEndSepPunct{\mcitedefaultmidpunct}
{\mcitedefaultendpunct}{\mcitedefaultseppunct}\relax
\EndOfBibitem
\bibitem[Asthana \latin{et~al.}(2023)Asthana, Kumar, Abraham, Grimsley, Zhang,
  Cincio, Tretiak, Dub, Economou, Barnes, and Mayhall]{Asthana_Mayhall:2023}
Asthana,~A.; Kumar,~A.; Abraham,~V.; Grimsley,~H.; Zhang,~Y.; Cincio,~L.;
  Tretiak,~S.; Dub,~P.~A.; Economou,~S.~E.; Barnes,~E.; Mayhall,~N.~J. Quantum
  self-consistent equation-of-motion method for computing molecular excitation
  energies{,} ionization potentials{,} and electron affinities on a quantum
  computer. \emph{Chem. Sci.} \textbf{2023}, \emph{14}, 2405--2418\relax
\mciteBstWouldAddEndPuncttrue
\mciteSetBstMidEndSepPunct{\mcitedefaultmidpunct}
{\mcitedefaultendpunct}{\mcitedefaultseppunct}\relax
\EndOfBibitem
\bibitem[Bravyi and Kitaev(2002)Bravyi, and Kitaev]{Bravyi_Kitaev:2002}
Bravyi,~S.~B.; Kitaev,~A.~Y. Fermionic Quantum Computation. \emph{Ann. Phys.}
  \textbf{2002}, \emph{298}, 210--226\relax
\mciteBstWouldAddEndPuncttrue
\mciteSetBstMidEndSepPunct{\mcitedefaultmidpunct}
{\mcitedefaultendpunct}{\mcitedefaultseppunct}\relax
\EndOfBibitem
\bibitem[Seeley \latin{et~al.}(2012)Seeley, Richard, and
  Love]{Seeley_Love:2012}
Seeley,~J.~T.; Richard,~M.~J.; Love,~P.~J. The {Bravyi-Kitaev} transformation
  for quantum computation of electronic structure. \emph{J.~Chem.\ Phys.}
  \textbf{2012}, \emph{137}, 224109\relax
\mciteBstWouldAddEndPuncttrue
\mciteSetBstMidEndSepPunct{\mcitedefaultmidpunct}
{\mcitedefaultendpunct}{\mcitedefaultseppunct}\relax
\EndOfBibitem
\bibitem[Kivlichan \latin{et~al.}(2018)Kivlichan, McClean, Wiebe, Gidney,
  Aspuru-Guzik, Chan, and Babbush]{Kivlichan_Babbush:2018}
Kivlichan,~I.~D.; McClean,~J.; Wiebe,~N.; Gidney,~C.; Aspuru-Guzik,~A.;
  Chan,~G. K.-L.; Babbush,~R. Quantum Simulation of Electronic Structure with
  Linear Depth and Connectivity. \emph{Phys. Rev. Lett.} \textbf{2018},
  \emph{120}, 110501\relax
\mciteBstWouldAddEndPuncttrue
\mciteSetBstMidEndSepPunct{\mcitedefaultmidpunct}
{\mcitedefaultendpunct}{\mcitedefaultseppunct}\relax
\EndOfBibitem
\bibitem[Yordanov \latin{et~al.}(2020)Yordanov, Arvidsson-Shukur, and
  Barnes]{Yordanov_Crispin:2020}
Yordanov,~Y.~S.; Arvidsson-Shukur,~D. R.~M.; Barnes,~C. H.~W. Efficient quantum
  circuits for quantum computational chemistry. \emph{Phys. Rev. A}
  \textbf{2020}, \emph{102}, 062612\relax
\mciteBstWouldAddEndPuncttrue
\mciteSetBstMidEndSepPunct{\mcitedefaultmidpunct}
{\mcitedefaultendpunct}{\mcitedefaultseppunct}\relax
\EndOfBibitem
\bibitem[Shlosberg \latin{et~al.}(2021)Shlosberg, Jena, Mukhopadhyay, Haase,
  Leditzky, and Dellantonio]{Shlosberg_Jena:2021}
Shlosberg,~A.; Jena,~A.~J.; Mukhopadhyay,~P.; Haase,~J.~F.; Leditzky,~F.;
  Dellantonio,~L. Adaptive estimation of quantum observables.
  \emph{arXiv:2110.15339} \textbf{2021}, \relax
\mciteBstWouldAddEndPunctfalse
\mciteSetBstMidEndSepPunct{\mcitedefaultmidpunct}
{}{\mcitedefaultseppunct}\relax
\EndOfBibitem
\bibitem[Gottesman(1998)]{Gottesman:1998}
Gottesman,~D. The {Heisenberg} representation of quantum computers.
  \emph{arXiv:quant-ph/9807006} \textbf{1998}, \relax
\mciteBstWouldAddEndPunctfalse
\mciteSetBstMidEndSepPunct{\mcitedefaultmidpunct}
{}{\mcitedefaultseppunct}\relax
\EndOfBibitem
\bibitem[Aaronson and Gottesman(2004)Aaronson, and
  Gottesman]{Aaronson_Gottesman:2004}
Aaronson,~S.; Gottesman,~D. Improved simulation of stabilizer circuits.
  \emph{Phys.~Rev.~A} \textbf{2004}, \emph{70}, 052328\relax
\mciteBstWouldAddEndPuncttrue
\mciteSetBstMidEndSepPunct{\mcitedefaultmidpunct}
{\mcitedefaultendpunct}{\mcitedefaultseppunct}\relax
\EndOfBibitem
\bibitem[Nielsen and Chuang(2000)Nielsen, and Chuang]{book_Nielsen_Chuang}
Nielsen,~M.~A.; Chuang,~I.~L. \emph{Quantum Computation and Quantum
  Information}; Cambridge University Press, 2000\relax
\mciteBstWouldAddEndPuncttrue
\mciteSetBstMidEndSepPunct{\mcitedefaultmidpunct}
{\mcitedefaultendpunct}{\mcitedefaultseppunct}\relax
\EndOfBibitem
\bibitem[Bansingh \latin{et~al.}(2022)Bansingh, Yen, Johnson, and
  Izmaylov]{Bansingh_Izmaylov:2022}
Bansingh,~Z.~P.; Yen,~T.-C.; Johnson,~P.~D.; Izmaylov,~A.~F. Fidelity Overhead
  for Nonlocal Measurements in Variational Quantum Algorithms. \emph{J.~Phys.\
  Chem.~A} \textbf{2022}, \emph{126}, 7007--7012\relax
\mciteBstWouldAddEndPuncttrue
\mciteSetBstMidEndSepPunct{\mcitedefaultmidpunct}
{\mcitedefaultendpunct}{\mcitedefaultseppunct}\relax
\EndOfBibitem
\end{mcitethebibliography}

\end{document}